\documentclass[aps,prb,twocolumn,showpacs,showkeys,groupedaddress]{revtex4}  
\usepackage{graphicx}  
\usepackage{dcolumn}   
\usepackage{bm}        
\usepackage{amssymb}   
\usepackage[ansinew]{inputenc}
\usepackage{hyperref}
\begin{document}


\title{Interplay of fractional quantum Hall states and localization in quantum point contacts}

\author{S.~Baer \footnote{Author to whom any correspondence should be addressed}, C.~R\"ossler, E.C.~de~Wiljes, P.-L.~Ardelt, T.~Ihn, K.~Ensslin, C.~Reichl, W.~Wegscheider} 
\affiliation{Solid State Physics Laboratory, ETH Z\"urich, 8093 Z\"urich, Switzerland}
\email{sbaer@phys.ethz.ch}

\date{\today}


\begin{abstract}
We investigate integer and fractional quantum Hall states in quantum point contacts (QPCs) of different geometries, defined in AlGaAs/GaAs heterostructures employing different doping and screening techniques. 
We find that, even in the highest mobility samples, interference and localization strongly influence the transport properties. We propose microscopic models for these effects, based on single- and many-electron physics.\\
For integer quantum Hall states, transport is modulated due to the self-consistent formation of compressible regions of enhanced or reduced density in the incompressible region of the constriction. 
In the fractional quantum Hall regime, we observe the localization of fractionally charged quasi-particles in the constriction and an interplay of single- and many-electron physics. At low electron densities and in comparatively weak magnetic fields, single-electron interference dominates transport.
Utilizing optimized growth and gating techniques, the $\nu$=5/2 state can be observed in a QPC, conserving the bulk properties in an unprecedented quality.
Our results might improve the understanding of the influence of localization on the transmission properties of QPCs, which is necessary for the interpretation of interference experiments employing QPCs, especially at $\nu=5/2$. 
\end{abstract}

\maketitle

\section{Introduction}

Localization plays a crucial role for understanding the exact quantization of the quantum Hall effect. 
In strong magnetic fields, electronic transport can be described in terms of narrow edge channels, leading to the well-known observations of $R_\mathrm{{xx}}\approx0$ and $R_\mathrm{{xy}}=h/(e^2\nu_\mathrm{bulk})$. 
In this case, varying the magnetic field only (de)populates localized states in the bulk, which do not affect transport because backscattering of the chiral edge states across the wide bulk region is negligible.
Alternative pictures, where the current is believed to flow in the bulk, exist (see for example Ref. \onlinecite{komiyama_edge_1998} for an overview). Also in this case, the (de)population of localized states plays the key role for the conductance quantization.
These theoretically predicted localized states have been investigated in various experiments using spatially resolved imaging techniques \cite{tessmer_subsurface_1998,ilani_microscopic_2004, martin_localization_2004, steele_imaging_2005,hashimoto_quantum_2008} or single-electron transistors fabricated on top of a two-dimensional electron gas (2DEG) \cite{wei_edge_1998}.
In a pioneering work by \textit{Ilani et al.} \cite{ilani_microscopic_2004}, the equilibrium properties of such localized states have been investigated in the bulk of a 2DEG with scanning single-electron transistor (SET) techniques.
The behavior of the bulk localizations was shown to be dominated by Coulomb blockade physics. 
This picture cannot be explained as a single-electron effect, but requires the formation of compressible and incompressible regions in the bulk, in analogy to an edge state picture that takes self-consistent screening into account \cite{chklovskii_electrostatics_1992, efros_homogeneous_1992, cooper_coulomb_1993}.
Here, the system is decomposed into compressible regions, in which potential fluctuations are screened and the density varies, and incompressible regions of constant density but varying background potential.
Apart from the aforementioned experiments which probe localizations on a very local scale in the bulk, conductance fluctuations in the quantum Hall regime, believed to be related to localized states, have been studied in direct transport experiments \cite{simmons_resistance_1989,simmons_resistance_1991,cobden_fluctuations_1999,machida_resistance_2001,couturaud_local_2009,peled_near-perfect_2003,staring_coulomb-blockade_1992,main_resistance_1994,timp_quantum_1987}. 
They have been investigated for example in Si-MOSFETs \cite{cobden_fluctuations_1999}, Graphene \cite{velasco_probing_2010,martin_nature_2009,branchaud_transport_2010}, InGaAs quantum wells \cite{granger_few-electron_2011} and in narrow AlGaAs/GaAs heterostructures \cite{simmons_resistance_1989,simmons_resistance_1991}, where localized states couple to the edge and thus become accessible. In the latter experiments \cite{simmons_resistance_1989,simmons_resistance_1991}, resistance fluctuations have been interpreted as magnetically bound states. As pointed out later \cite{lee_comment_1990,rosenow_influence_2007,halperin_theory_2011}, Coulomb blockade effects are of great importance for such experiments and have to be taken into account for the interpretation of $B$-field and gate-voltage periodicities.
In the work of \textit{Cobden et al.} \cite{cobden_fluctuations_1999}, conductance fluctuations in the quantum Hall regime span a distinct pattern in the density vs. magnetic field plane, with resonances parallel to neighboring conductance plateaus, similar to the phase diagram obtained by \textit{Ilani et al.} 
This has been interpreted as Coulomb charging of localized states in the bulk of the employed small structures. The absence of a clear periodicity suggests either the contribution of many localized states or the validity of other interpretations\cite{machida_resistance_2001}, which are based on the presence of a network of compressible stripes.\\
More recently, scanning gate experiments have tried to combine spatial resolution with transport \cite{hackens_imaging_2010,martins_scanning_2013,martins_coherent_2013}. \textit{Hackens et al.} have investigated Coulomb dominated islands inside quantum Hall interferometers\cite{hackens_imaging_2010}. Modulations of transport, due to the coupling of the localized islands to the edge states were found. In contrast to this behavior dominated by Coulomb charging, recent experiments \cite{martins_coherent_2013} report phase coherent tunneling across constrictions in the quantum Hall regime.\\
Quantum point contacts are one of the conceptually most simple, though interesting systems studied in mesoscopic physics. Despite their simplicity, complex many-particle phenomena like g-factor enhancement \cite{martin_enhanced_2008,yoon_nonlocal_2009,rossler_transport_2011} or the 0.7 anomaly \cite{thomas_possible_1996,kristensen_bias_2000,cronenwett_low-temperature_2002,komijani_evidence_2010,micolich_what_2011,komijani_origins_2013} are observed. Furthermore, they offer the possibility of locally probing a physical system, which is for example used in charge detection experiments \cite{gustavsson_counting_2006-1,gustavsson_frequency-selective_2007,rossler_tunable_2013,baer_cyclic_2013} or tunneling experiments in the quantum Hall regime \cite{milliken_indications_1996,chang_observation_1996,roddaro_interedge_2004}.
This allows us to employ QPCs in the quantum Hall regime for investigating the influence of disorder-induced localizations on transport. The influence of localized states on (fractional) quantum Hall states confined in QPCs is not fully understood. 
Furthermore, the influence of individual localizations on non-equilibrium transport was not accessible in the mentioned transport experiments. In contrast, scanning SET experiments provided information about individual localizations, but not about their influence on transport.
For the interpretation of interference experiments in the quantum Hall regime \cite{zhang_distinct_2009,kou_coulomb_2012,camino_quantum_2007,mcclure_fabry-perot_2012,willett_measurement_2009}, a detailed understanding of the transmission properties of single QPCs is necessary. 
Even in 2DEGs with the highest mobilities technologically achievable at the moment, disorder significantly influences transport through the QPC, as soon as a perpendicular magnetic field is applied. 
We show that even in simple QPCs complicated behavior can be observed, which is interpreted in terms of single- and many-electron physics of individual disorder-induced localizations.
We argue that the influence of localizations can be minimized by employing growth and gating techniques, which result in a very steep QPC confinement potential (perpendicular to transport direction) and low disorder in the channel. By this, the $\nu=5/2$ state can be confined to a QPC without noticeable backscattering and preserving the bulk properties in an unprecedented quality, giving a good starting position for tunneling- and interference experiments in the second Landau level.\\

\section{Experimental details}
The QPCs used in this study are defined by electron-beam lithography and subsequent Ti/Au evaporation on photo-lithographically patterned high-mobility wafers. Constrictions with different geometries have been studied for this manuscript (see Tab.\ref{TabelleSamples} for an overview). 
For the 250 nm wide QPC I.a and the 500 nm wide QPC I.b, a 30 nm wide quantum well with a carrier sheet density $n_\mathrm{s}\approx3.04\times 10^{11}~$cm$^{-2}$ and a mobility $\mu\approx13\times 10^6~$cm$^{2}$/Vs has been used. 
In this structure the 165 nm deep quantum well is neighbored by two $\delta$-Si doped GaAs layers, enclosed in 2 nm thick layers of AlAs. These screening layers reside 70 nm below and above the 2DEG. The electrons in the AlAs wells populate the X-band and provide additional low-mobility electron layers, which screen the $\Gamma$-electrons in the 2DEG from remote ionized impurities. 
Due to the screening layers, hysteretic and time-dependent processes make gating difficult. The gating properties of these wafers have been studied earlier \cite{rossler_gating_2010}.
The 1.2 $\mathrm{\mu}$m wide QPC III.a has been fabricated on a wafer which employs a similar growth technique ($\mu \approx 17.8\times 10^6~$cm$^2$/Vs, $n_\mathrm{s} \approx 2.13\times 10^{11}~$cm$^{-2}$, 250 nm deep, 30 nm wide quantum well, screening layers 100 nm below and above the 2DEG). The high mobility structures used for QPCs I.a,b and III.a are optimized for the $\nu=5/2$ state without the requirement of prior LED illumination \cite{reichl_increasing_2014}. 
QPC II.a, QPC II.b (both 700 nm wide), QPC II.c (900 nm wide) and QPC II.d (800 nm wide) were fabricated on a single side doped GaAs/Al$_\mathrm{x}$Ga$_\mathrm{1-x}$As heterostructure with a mobility of approximately $8\times10^6$ cm$^2$/Vs and a 320 nm deep 2DEG with an electron density of approximately $1.5\times10^{11}$cm$^{-2}$. Hysteresis effects are much less pronounced in these structures which employ a reduced proportion of Al in the spacer layer between the doping plane and the 2DEG (x=0.24 compared to typically x=0.30$-$0.33). 
Long-range scattering is therefore reduced, thus facilitating the formation of the $\nu$=5/2 state and other fragile FQH states \cite{pan_impact_2011,gamez_5/2_2013}.
The measurements have been conducted in a dilution refrigerator at a base temperature of approximately 85 mK and in magnetic fields up to 13 T. Measurements of QPC III.a have been performed in a dry dilution refrigerator with an electronic temperature of approximately 12 mK, achieved by massive filtering and thermal anchoring at every temperature stage. Standard four-terminal lock-in measurement techniques have been used to measure $R_\mathrm{xx}$ and $R_\mathrm{xy}$ of the bulk 2DEG and the differential conductance the QPC, $G=\partial I_\mathrm{AC}/\partial V_\mathrm{diag}$, which gives access to the effective QPC filling factor $\nu_\mathrm{QPC}$\cite{beenakker_quantum_2004}. Here, the voltage drop $V_\mathrm{diag}$ is measured diagonally across the QPC. 

\begin{table*}
\begin{tabular}{ccccc}
 \toprule
QPC 		& $w$ (nm) 	& Heterostructure																& $n_s$ (10$^{11}$ cm$^{-2}$) 	& $\mu$  (10$^{6}$ cm$^{2}$/Vs)	\\ \toprule
QPCI.a 	& 250 			 & 30 nm QW, $\delta$-Si screening						 		& 3.0	 											& 13.0 												\\ \cline{1-2}
QPCI.b 	& 500 			& 																					&  													& 														\\\hline
QPCII.a 	& 700 			&  																					& 1.5												& 8.0 													\\\cline{1-2}
QPCII.b	& 700 			& Single-side doped 														& 													& 														\\\cline{1-2}
QPCII.c 	& 900 			& GaAs/Al$_{0.24}$Ga$_{0.76}$As heterostructure 	& 													& 														\\\cline{1-2}
QPCII.d 	& 800 			&  																					&  													& 														\\ \hline
QPCIII.a 	& 1200 		& 30 nm QW, $\delta$-Si screening 								& 2.1	 											& 17.8 												\\ \hline
\end{tabular}
\caption{Overview of the different samples used in this study. Channel width $w$, electron sheet density $n_s$ and mobility $\mu$ are indicated for the different QPCs.}
\label{TabelleSamples}
\end{table*}

\section{Results and discussion}
The main part of this article will be organized as follows: First, an exemplary quantum Hall phase diagram will be discussed (\ref{phase}). The influence of different QPC geometries on the width of the incompressible region separating the edge states and the density distribution is discussed in section \ref{geometry}. In the main part of our article, section \ref{main}, QPC resonances are characterized and explained via a microscopic model. A short summary of this central chapter is given afterwards. The resonances' dependence on the spatial position of the conducting channel inside the QPC is investigated in the following (\ref{AsymmetrieChap}). At the end, methods for confining the most fragile fractional quantum Hall states are discussed (\ref{Fragile}).
\subsection{Quantum Hall phase diagram of a QPC}
\label{phase}
\begin{figure}
\begin{center}
\includegraphics[width=8cm]{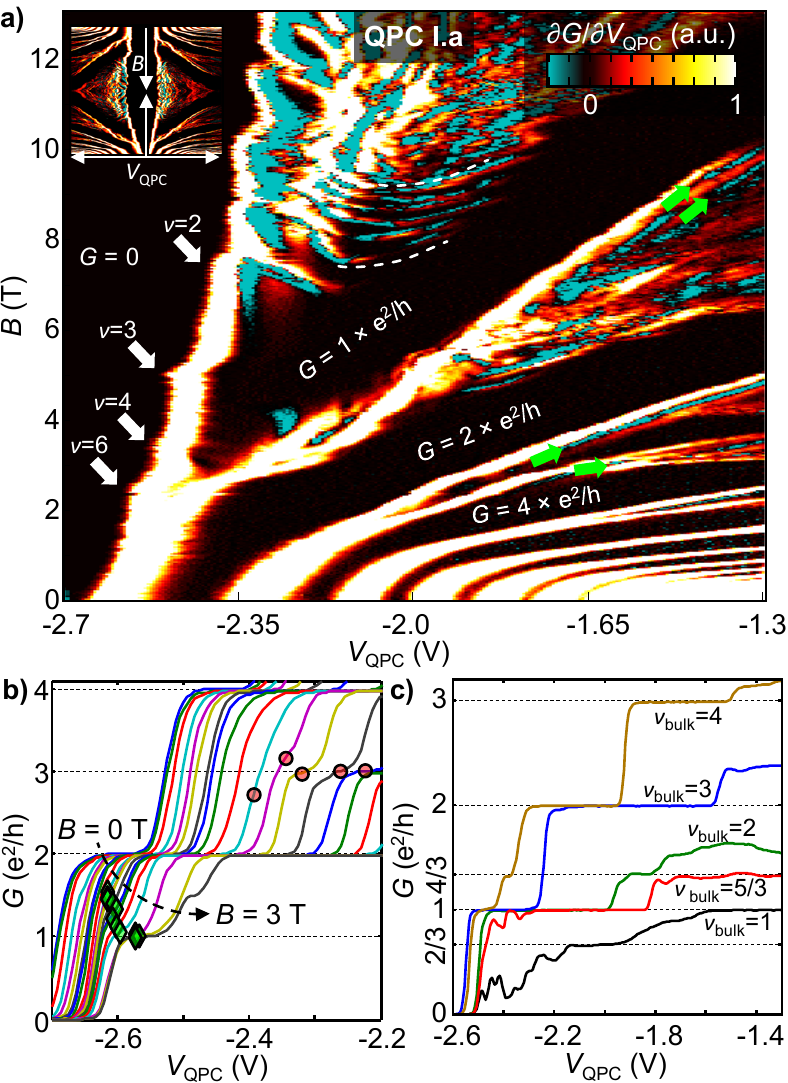}
\end{center}
\caption{(Color online) \textbf{a:} Transconductance of QPC I.a as a function of the QPC voltage and the magnetic field. Conductance plateaus at multiples of $e^2/h$ can be seen as black regions. Resonances, bending in the $B-V_\mathrm{QPC}$-plane are indicated by white dashed lines. The 1/$B$ periodic kinks (white arrows) are believed to originate from a change of the filling factor in the bulk. A possible combination of bulk filling factors is indicated. Inset: full B-field and voltage dependence of the system. Here, the voltage was swept from -1.3 V $\rightarrow$ -2.7 V $\rightarrow$  -1.3 V repeatedly and the $B$-field was stepped from 0 T $\rightarrow$ 13 T $\rightarrow$  0 T. \textbf{b:} The $B$=0 QPC conductance plateaus at multiples of $2\times e^2/h$ spin-split for increasing magnetic fields (magnetic field from 0 T to 3 T). For $0<G<2\times e^2/h$ and $2\times e^2/h<G<4\times e^2/h$, local minima in the slope of the conductance are marked by green diamonds or red circles. In contrast to the second and third subband, the spin-splitting of the lowest subband starts at conductance values of approx. $0.7\times2\times e^2/h$ and approaches $1\times e^2/h$ as the magnetic field strength is increased \cite{wharam_one-dimensional_1988, van_wees_quantized_1988-1}. \textbf{c:} At strong magnetic fields (with bulk filling factors $\nu_{bulk}$), conductance plateaus corresponding to different integer and fractional filling factors in the QPC can be observed.}
\label{Schmetterling}
\end{figure}

Fig. \ref{Schmetterling}.a shows the differential conductance $G$ (plotted: numerical derivative $\partial G/\partial V_\mathrm{QPC}$ in colorscale) of QPC I.a as a function of the voltage applied to the QPC gates ($V_\mathrm{QPC}$) and a perpendicular magnetic field $B$. 
At zero magnetic field, the well-known QPC conductance quantization in multiples of $2\times e^2/h$ is found. 
As the magnetic field is increased, conductance steps (or plateaus), seen as maxima (or black areas) of $\partial G/\partial V_\mathrm{QPC}$, bend to more positive QPC voltages, due to magneto-electric depopulation of the QPC channel \cite{buttiker_quantized_1990,beenakker_quantum_2004}. 
The quantized conductance plateaus successively develop into regions of constant effective filling factor of the QPC ($\nu_\mathrm{QPC}$) with a diagonal resistance $R_\mathrm{diag}=h/(e^2\nu_\mathrm{QPC})$. 
In this regime the spin splitting is sufficiently strong to also observe conductance plateaus at $G=1,3,5,...~\times e^2/h$ \cite{wharam_one-dimensional_1988, van_wees_quantized_1988-1}. 
The low-field behavior of the spin-splitting is shown in Fig. \ref{Schmetterling}.b. 
Numerically extracted local minima of the slope of the conductance curve have been marked with (green) diamonds / (red) circles. As $B~\rightarrow~0$, the $G=1\times e^2/h$ plateau seems to join the $0.7\times2\times e^2/h$ anomaly\cite{graham_interaction_2003}. 
No similar behavior can be observed at $G=3\times e^2/h$ and $G=5\times e^2/h$. 
In the quantum Hall regime, conductance curves of the QPC (Fig \ref{Schmetterling}.c) show fractional effective filling factors at $\nu_\mathrm{QPC}=2/3$ and $\nu_\mathrm{QPC}=4/3$ for different integer and fractional filling factors of the bulk ($\nu_\mathrm{bulk})$.
The shape of the boundary of the $G=0$ region of Fig. \ref{Schmetterling}.a is determined by different effects: first, an increasing magnetic field leads to magneto-electric depopulation due to an increase of the single-particle electron energy, thus moving the pinch-off region to less negative gate voltages. 
In addition, time-dependent and hysteretic processes of the X-electron screening layers lead to an additional drift of the pinch-off line towards less negative QPC voltages. The small inset shows the $B$-field and voltage dependence of the system. Here, the voltage was swept from -1.3 V $\rightarrow$ -2.7 V $\rightarrow$  -1.3 V repeatedly (horizontal axis) and the $B$-field was stepped from 0 T $\rightarrow$ 13 T $\rightarrow$  0 T (vertical axis). Upper and lower part of the inset are not mirror symmetric - over the time of the measurement, the pinch-off-line drifts towards less negative voltages, indicating a time-dependence of the system. 
Furthermore, changes of the filling factor in the bulk can lead to an abrupt decrease of the Fermi energy of the system as observed in quantum dots (QDs) \cite{ciorga_addition_2000}. This effect is believed to cause the $1/B$-periodic kinks in the pinch-off line. When increasing the $B$-field across the kinks, the pinch-off line suddenly moves towards more positive QPC voltages (marked by white arrows in Fig. \ref{Schmetterling}.a), though an assignment to the individual filling factors in the bulk is not uniquely possible, probably due to a reduced density in the bulk near the QPC, which governs the local coupling of bulk states into the QPC.

In the regions of Fig. \ref{Schmetterling}.a where the QPC filling factor is changing, the QPC conductance does not vary monotonically. 
For $G>1\times e^2/h$, resonances which are parallel to the boundary of one of the neighboring conductance plateaus are observed (green (gray) arrows). 
In contrast, the region $G<1\times e^2/h$ shows resonant features without any preferred slope, or even with varying slope at different $B$-fields (a set of bending resonances is indicated by the white dashed line). 
Very similar resonances have been found in several QPCs in different cooldowns.
The origin of these resonances will be discussed later in the framework of single- and many-electron physics.

\subsection{Influence of QPC geometry on incompressible separating region and density distribution}
\label{geometry}
\begin{figure}
\begin{center}
\includegraphics[width=8cm]{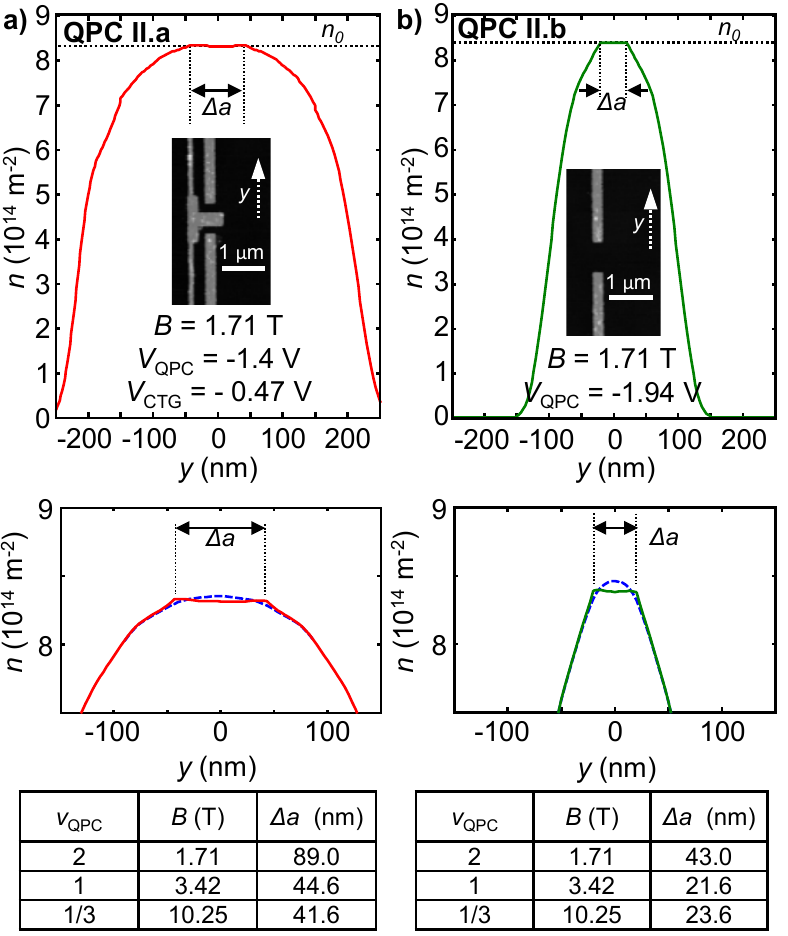}
\end{center}
\caption{(Color online) Density distributions in the channels of QPCII.a (\textbf{a}) and QPCII.b (\textbf{b}). The original zero magnetic field density (dashed (blue), blow-up in second row) has been altered by the formation of an incompressible region in the center of the constriction at $B=1.71$ T ($\nu_\mathrm{QPC}=2$), resulting in a region of constant density $n_0$. The width $\Delta$a of the incompressible region is indicated for different filling factors $\nu_\mathrm{QPC}$ in both QPCs.}
\label{Simulation}
\end{figure}

To be able to understand the mechanisms behind the resonances in more detail, we have investigated two different QPC designs, fabricated on a 2DEG of lower density. 
QPC II.a is 700 nm wide with a top-gate above the conducting channel (see inset Fig. \ref{Simulation}.a). QPCs II.b  (inset Fig. \ref{Simulation}.b) and II.c are standard 700 nm / 800 nm wide QPCs.
Density profiles in the y-direction (along the lateral confinement potential) for the two QPC designs (at $B$ = 0) have been obtained from a self-consistent bandstructure calculation using next\textbf{nano}. 
The doping concentration was adjusted to account for surface charges and to reproduce the gate pinch-off voltages correctly. 
The applied voltages to the gates were chosen such that the calculated density at $B$ = 0 corresponds to the density necessary for $\nu_\mathrm{QPC}$ = 2 at $B$ = 1.71 T, as in the measurements.
Using the calculated density profile at $B$ = 0, we have calculated the altered density profile when a compressible region is formed in the center of the QPC (Fig. \ref{Simulation}) and the width $\Delta$a of the incompressible region for different QPC filling factors, using the electrostatic model of \textit{Chklovskii et al.}\cite{chklovskii_ballistic_1993}. 
In this model, perfect metallic screening in the compressible regions is assumed. For the gap energies, $\hbar \omega_c$ and $\tilde{g}\mu_BB$ with an exchange enhanced $\tilde{g}\approx 4~$\cite{rossler_transport_2011} have been used as estimates for $\nu=2$ and $\nu=1$. The energy gap at $\nu=1/3$ has been measured (see appendix).
In Fig. \ref{Simulation}, the resulting self-consistent densities (for $\nu_\mathrm{QPC}$ = 2) are shown as solid lines. The original density at $B$ = 0 (dashed (blue) line, second row) is modified by the formation of an incompressible stripe (constant density $n_0$) in the center of the channel.
When a negative voltage is applied to the channel top-gate (CTG) of QPC II.a, the subband minimum is lifted. 
In this situation, the curvature of the electron density in the center of the constriction is small.
When QPC II.b is tuned to a similar density in the constriction, the density curvature in the center is much greater, leading to a narrower compressible region (Fig.\ref{Simulation}.b). 
Comparing $\Delta$a of the two QPCs, we conclude that for QPC II.a, a significantly wider incompressible region is expected according to the model of \textit{Chklovskii et al.} \cite{chklovskii_ballistic_1993}. The widths $\Delta$a range from approximately 20 nm to 90 nm.
Disorder potential fluctuations have typical length scales of the order of 100 nm \cite{sohrmann_compressibility_2007,ilani_microscopic_2004}.
If the amplitude of such a disorder potential fluctuation in the incompressible region in the center of the QPC is large enough to create an intersection of the Landau level with the Fermi energy (Fig. \ref{Schema}.B,C, left column), compressible regions of enhanced or reduced density (Fig. \ref{Schema}.B,C, middle column) are formed.
Thus, the small width of the incompressible region in QPCII.b (and hence in the QPCII.c with similar geometry) makes it less likely that a disorder potential fluctuation leads to the formation of a localization in the constriction. 
Furthermore, the coupling to such a localization is strongly varied as the width of the separating incompressible region changes, making the observation of periodic charging of a single localization impossible. 
To observe periodicities and study resonances in more detail, we now investigate electronic transport in QPC II.a, where a much stronger influence of disorder-induced localizations is expected. Here, a periodic behavior is expected over a larger parameter range, as the width of the incompressible regions separating edge and localizations is sufficiently wide.

\subsection{Characterization of QPC resonances and microscopic model}
\label{main}
\subsubsection{Periodic conductance oscillations in QPCs of different geometries}
\begin{figure*}
\begin{center}
\includegraphics[width=16cm]{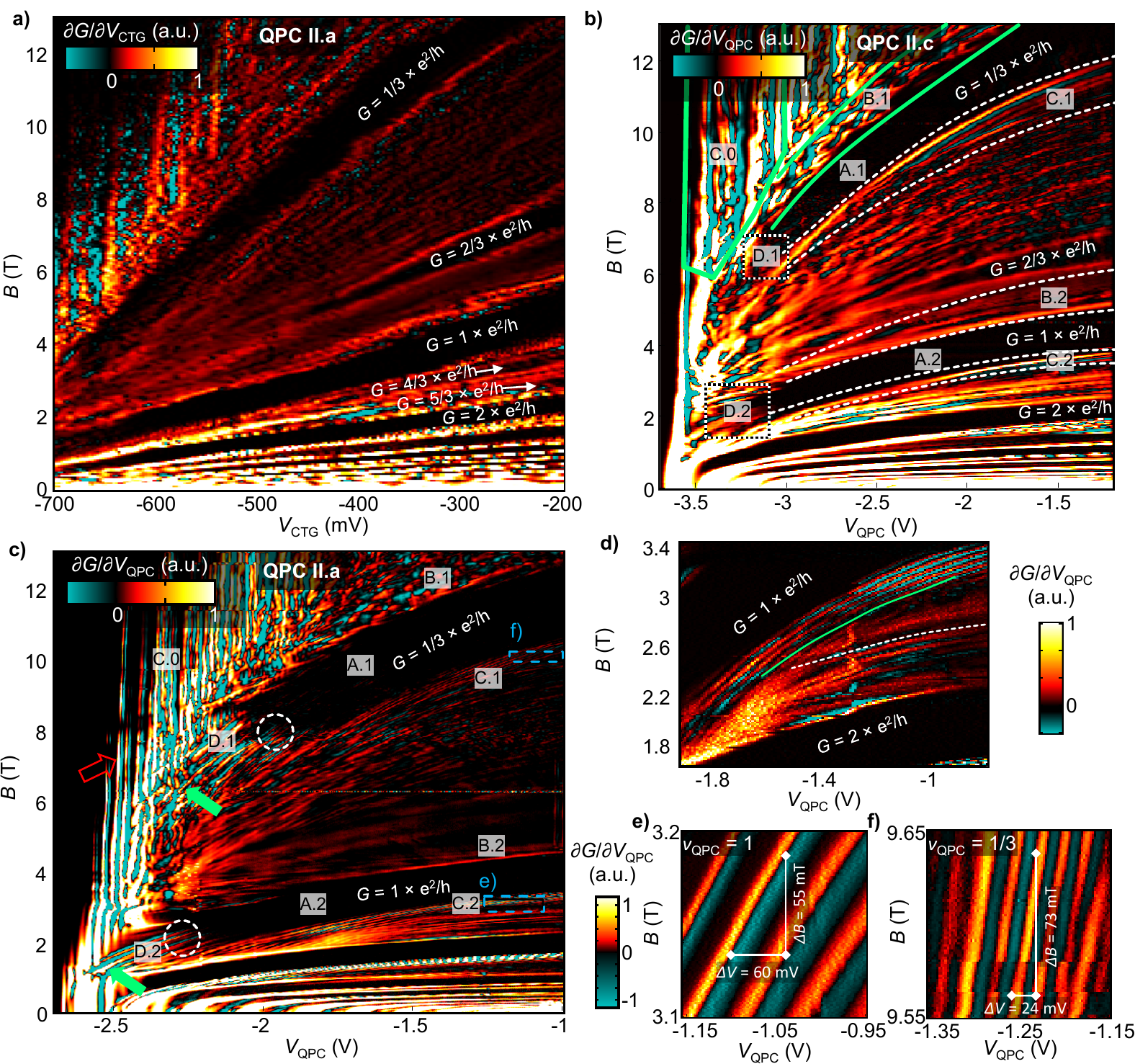}
\end{center}
\caption{(Color online) \textbf{a}: Transconductance of QPC II.a as a function of the voltage $V_{\mathrm{CTG}}$ and magnetic field $B$. Here, the density is tuned roughly linearly by the gate voltage. \textbf{b}: 
Transconductance of QPC II.c as a function of $V_\mathrm{QPC}$. Pronounced fractional and integer filling factors are observed (black regions A.1, A.2, etc.). 
Apart from these regions of nearly-perfect transmission, disorder modulates transport in other regions: for small transmission (C.0, C.1), small backscattering (B.1) and at the low density, low $B$-field end of the conductance plateaus (D.1). 
A similar behavior is found when one underlying edge state is perfectly transmitted (regions B.2 - D.2). 
\textbf{c:} Transconductance of QPC II.a, when -400 mV are applied to the CTG. \textbf{d:} Zoom of Fig. \ref{Faecher}.c: transition from $\nu_\mathrm{QPC}=2$ to $\nu_\mathrm{QPC}=1$. 
Two distinct slopes (green solid / white dashed lines), parallel to the boundary of the neighboring conductance plateaus, are observed. 
\textbf{e,f:} Close-ups of the conductance oscillations for $\nu_\mathrm{QPC}=1$ and $\nu_\mathrm{QPC}=1/3$ (enframed areas in \textbf{c}).}
\label{Faecher}
\end{figure*}
The filling factor spectra of QPC II.a and QPC II.c are investigated similarly to the measurement of Fig. \ref{Schmetterling}.a, by varying the QPC gate voltage versus the magnetic field $B$. 
First, the channel top-gate voltage $V_\mathrm{CTG}$ has been varied (Fig. \ref{Faecher}.a). 
This gate varies the density of the channel roughly linearly with applied voltage (neglecting filling-factor dependent capacitances), as seen from the slope $dB/dV_\mathrm{CTG}\varpropto 1/\nu_\mathrm{QPC}$ of the conductance plateaus, which show up as black areas of quantized conductance. 
In addition to the full series of integer filling factors, fractional states at $\nu_\mathrm{QPC}=1/3$, 2/3, 4/3 and 5/3 can be observed. 
Close to the low- and high density edges of the conductance plateaus, sets of conductance oscillations with a slope parallel to the boundaries are observed, similar to the ones observed in small Hall-bars \cite{cobden_fluctuations_1999,machida_resistance_2001}. 
The slope and number of these resonances are independent of density and magnetic field strength.\\
A qualitatively similar behavior can be found for QPC II.c as $V_\mathrm{QPC}$ is varied (Fig.\ref{Faecher}.b). 
Here, regions of perfect transmission have been marked (A.1). 
Modulations occur at the low-density side (B.1) and high-density side of the conductance plateaus (C.1) or pinch-off (C.0). 
Furthermore, resonances at the low-density and low-$B$-field end of conductance plateaus are observed (D.1). 
These resonances disappear as the density and $B$-field strength increase. 
Similar regions can be attributed to higher filling factors, for which underlying edge states are perfectly transmitted (A.2-D.2).
\\
As mentioned above, a much stronger influence of a disorder-induced localization is expected for QPC II.a, as the wider incompressible region is much more likely to accommodate one or several extrema of the disorder potential. 
Here, quasi-periodic conductance modulations should occur over a larger parameter range, as the width of the incompressible regions separating edge and localizations is sufficiently wide.
In order to verify this expectation, Fig. \ref{Faecher}.c shows the transconductance of QPC II.a, obtained by keeping $V_\mathrm{CTG}$ fixed while varying $V_\mathrm{QPC}$ and $B$. 
Compared to Fig. \ref{Faecher}.b, more pronounced conductance oscillations are observed (red empty arrow). 
Especially gate voltage regions close to pinch-off are now dominated by equidistant conductance peaks parallel to the magnetic field axis. 
Regions C and D overlap, which can be seen from the coexistence of two different distinguishable slopes (indicated by green solid arrows). 
In the integer quantum Hall regime (for example in Fig. \ref{Faecher}.d), conductance oscillations with distinct slopes are observed between neighboring conductance plateaus.
The resonances are parallel to either of the two neighboring plateau boundaries (Fig. \ref{Faecher}.d, green solid / white dashed line). 
Close to $\nu_\mathrm{QPC}=1/3$, strong resonances, parallel to the conductance plateau occur (Fig. \ref{Faecher}.f). At lower B-Fields, between $\nu_\mathrm{QPC}=1/3$ and $\nu_\mathrm{QPC}=1$, weak modulations with an intermediate slope are observed. 
\begin{figure*}
\begin{center}
\includegraphics[width=16cm]{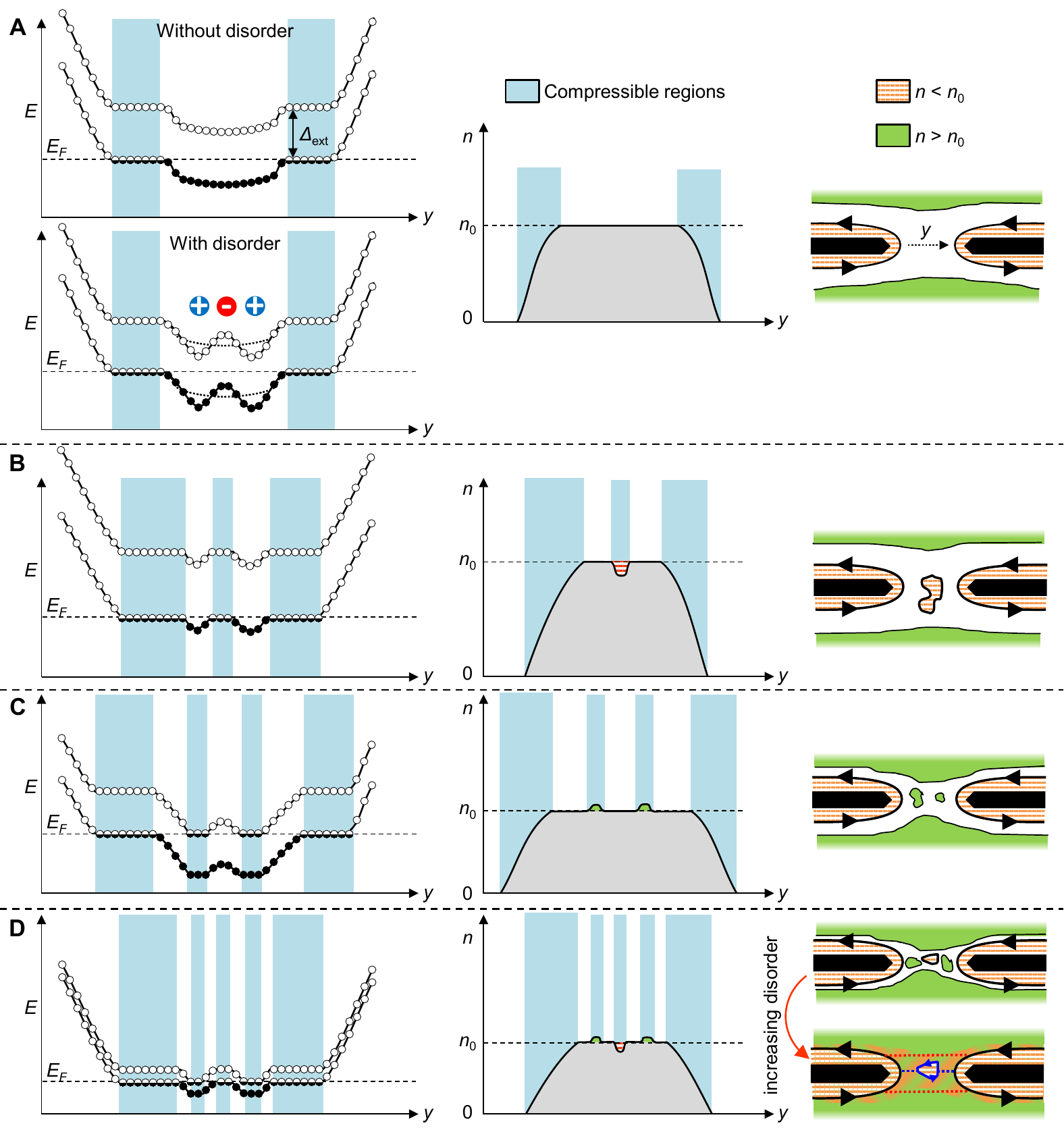}
\end{center}
\caption{(Color online) Schematic guiding center energies of extended states and densities within a QPC for the different transmission situations A-D, as indicated in Fig. \ref{Faecher}.b. Empty / filled circles symbolize empty or occupied states.
In the case of perfect transmission (\textbf{A}), adding an exemplary disorder potential does not alter the density distribution within the channel. As density fluctuations in the transmitted (\textbf{B}) or energetically lowest reflected (\textbf{C}) Landau level lead to partially occupied states at the Fermi energy, compressible regions of enhanced (\textbf{C}, green (dark grey) area) or reduced density (\textbf{B}, red striped area) are formed within the incompressible region. 
This gives rise to a quantized charge on the compressible regions of enhanced or reduced density formed in the incompressible region which separates the edge states. For smaller magnetic fields or stronger disorder fluctuations, wide compressible regions are absent and only states  below the Fermi energy are occupied. Here, compressible regions of enhanced and reduced density modulate the transport in the constriction at the same time (\textbf{D}). As wide incompressible regions are absent, the compressible regions of enhanced or reduced density are no longer governed by Coulomb dominated physics. Here, single-electron resonances arise from localized states, encircling a certain number of magnetic flux quanta. In contrast to the Coulomb dominated mechanism, such single-electron resonances give rise to a dependence in the $B-V_\mathrm{QPC}$ plane which may differ from the slope of the conductance plateaus.}
\label{Schema}
\end{figure*}
\subsubsection{Screening and localization model}
The mechanism which gives rise to the different resonances in regions A-D can be understood in terms of an edge-state picture which takes non-linear screening of potential fluctuations into account (Fig. \ref{Schema}). Similar models have been employed to understand bulk localizations in scanning SET and scanning capacitance experiments \cite{ilani_microscopic_2004,martin_localization_2004,steele_imaging_2005,steele_imaging_2006}. 
Regions of locally enhanced or reduced density are formed on top of the background density, associated with different extended quantum Hall states. These localizations in the constriction couple to the edge states and give rise to conductance oscillations.
In Fig. \ref{Schema}, the guiding center energies of two extended quantum Hall states are shown as a function of the spatial direction $y$, intersecting the QPC channel (Fig. \ref{Schema}.A, left column). Empty / filled circles symbolize empty or occupied states.
The extended states could for example be associated with Landau levels (in this case $\Delta_\mathrm{ext}=\hbar \omega_c$), spin-split Landau levels ($\Delta_\mathrm{ext}=\tilde{g}\mu_B B$ with an exchange enhanced $\tilde{g}$), or $\Lambda$-levels of composite fermions, corresponding to a FQH state at $\nu$=1/$m$ \cite{macdonald_edge_1990,brey_edge_1994,sim_composite-fermion_1999}($\Delta_\mathrm{ext}=\Delta_{1/m}$ is the energy gap of the FQH state).
For simplicity, we will constrain the discussion in the following to the situation, where extended states arise from a Landau level splitting. 
If spin-split Landau levels or $\Lambda$-levels are considered, an analog picture can be constructed.\\
In Fig. \ref{Schema}.A, energies of the second Landau level are far above the Fermi energy. 
In the most simple edge state picture \cite{halperin_quantized_1982,buttiker_absence_1988}, Landau level energies are bent up by the confinement potential of the QPC, giving rise to chiral edge states at the intersections with the Fermi energy, thus leading to a step-wise density increase towards the bulk of the sample. 
However, self-consistency of the Poisson and Schr\"odinger equations at a smooth, electrostatically defined edge \cite{chklovskii_electrostatics_1992} leads to a screened potential and smooth density variations in compressible regions of finite width (\ref{Schema}.A, middle column).
In this compressible region, partially filled states (half-filled circles) reside at the Fermi energy. 
The density of electrons in the lowest Landau level is constrained via $n\le n_0=2 eB/h$, due to its finite degeneracy. 
Where the Landau level energy lies below the Fermi energy, all states are occupied (filled circles) and this maximum density is reached. 
Potential fluctuations can no longer be screened, in contrast to the ideally perfect screening in compressible regions where the potential is flat. 
In this picture, compressible regions in between regions of constant filling factors $\nu_1$ and $\nu_2$ contribute $G=e^2/h\times\Delta\nu$ to the conductance, where $\Delta\nu=\nu_2-\nu_1$ \cite{beenakker_theory_1991}. Alternate models exist, where the current is flowing in the bulk (see for example Ref. \onlinecite{komiyama_edge_1998} for an overview). In our case however, the details of the current distribution in the QPC are not important, as only the total conductance through the QPC can be measured. 
A schematic spatial density distribution within the QPC is shown in the right column of Fig. \ref{Schema}.A. 
Here, the boundaries between compressible and incompressible regions are indicated as black arrows. 
For simplicity, these will be referred to as 'edge states' from now on. 
In this picture, the edge state is perfectly transmitted through the QPC constriction (between black polygons) and both counter-propagating directions are separated by a wide incompressible region \cite{chklovskii_ballistic_1993} (white), yielding a quantized QPC conductance. 
Far away from the QPC, additional Landau levels eventually fall below the Fermi energy, leading to additional compressible regions where the density increases towards its bulk value (green (dark grey) area). 
Adding schematic potential fluctuations (\ref{Schema}.A, left column) does not change the overall situation, as long as no states in the second Landau level become occupied. 
This is the analog situation in the regions A.1 and A.2 of Fig. \ref{Faecher}.b.
As the magnetic field strength is increased, Landau levels are lifted in energy, leading to a narrower incompressible region in the center of the QPC between the edge states. 
The density is locally reduced (\ref{Schema}.B, middle column) where maxima of the potential fluctuations intersect the Fermi energy (\ref{Schema}.B, left column). 
This leads to the formation of a compressible region of reduced density (red striped) that is separated from the edge states via incompressible stripes.\\
For an increasing disorder amplitude or decreasing Landau level splitting, compressible regions of enhanced or reduced density can occur in the constriction at the same time, explaining the simultaneous visibility of resonances with a different slope in Fig. \ref{Faecher}.c (indicated by solid green arrows). 
When disorder dominates over the Landau level splitting, i.e. when the the gradient of the background potential $\partial V/\partial y$ becomes comparable to $E_\mathrm{gap}/l_\mathrm{B}$\cite{beenakker_edge_1990}, where $l_\mathrm{B}$ is the magnetic length, the system is no longer described by a many-electron picture with screening via compressible and incompressible regions (\ref{Schema}.D). 
In that confinement-dominated case, single-electron states localized around a potential minimum or maximum in the constriction enclose a fixed number of flux quanta \cite{jain_quantum_1988} (Fig. \ref{Schema}D, solid blue line). As the area of the localized state is tuned non-linearly with the QPC gate voltage, resonances with varying slope in the $B-V_\mathrm{QPC}$ plane are expected \cite{ilani_microscopic_2004}. \\
Alternate tunneling paths that lead to a qualitatively similar behavior have been proposed \cite{van_loosdrecht_aharonov-bohm_1988}. Here, non-adiabaticity of the QPC potential leads to enhanced tunneling between the edge channels at the entrance and exit of the constriction (Fig. \ref{Schema}D, red dashed lines).
In contrast, the situation of Fig. \ref{Schema}.B is described by Coulomb dominated physics of the compressible region of reduced density inside the constriction. 
Here, electron-electron interactions lead to a potential with compressible and incompressible regions. 
The charge of the compressible region of reduced density is quantized, leading to a certain slope in the $B-V_\mathrm{QPC}$ plane \cite{ilani_microscopic_2004}, whenever an electron is added or removed from the compressible region of reduced density. The slope is uniquely determined by the filling factor of the incompressible region in which the compressible region of enhanced or reduced density is formed and equals the slope of the corresponding conductance plateaus in the $V_\mathrm{QPC}$-$B$-field plane.
This explains why resonances only occur with one of the slopes of the neighboring conductance plateaus (Fig. \ref{Faecher}.d).
Conductance resonances are only visible in the transport data when the incompressible region between the edge states and the compressible region of reduced density is sufficiently small, allowing for resonant backscattering across the constriction. 
This is the case as the conductance starts to decrease below the plateau value, as in Fig. \ref{Faecher}.b B.1 and B.2.
Similarly, potential minima of the second Landau level fall below the Fermi energy, as the magnetic field strength is decreased (\ref{Schema}.C, left column), leading to compressible region of enhanced density within the incompressible region separating the edge states. As additional transmission sets in (Fig. \ref{Faecher}.b, C.1 and C.2), the coupling of these compressible region of enhanced density leads to a periodic modulation of the transmission.\\
In this discussion, the additional complication of possible edge reconstruction of integer quantum Hall (IQH) edge states \cite{venkatachalam_local_2012} has not been taken into account. Furthermore, we observe faint conductance plateaus at $G=2/3 \times e^2/h$ in the QPC. This state is clearly visible in the QPCII.c (Fig. \ref{Faecher}.b) and in QPCII.a when the voltage applied to the CTG is swept (Fig. \ref{Faecher}.a). Surprisingly, the $\nu$=2/3 state is not observed, when the QPC voltage of QPCII.a is swept while a constant voltage is applied to the CTG (Fig. \ref{Faecher}.c). 
The edge structure of the $\nu=2/3$ state is still not understood in detail. 
Theory and experimental findings suggest, that this state may consist of a $\delta\nu$=1 IQH edge state and a counter-propagating $\delta\nu$=-1/3 edge state of holes which are equilibrated by interaction, resulting in a single chiral charged mode and a counter-propagating neutral mode \cite{girvin_particle-hole_1984,macdonald_edge_1990,wen_electrodynamical_1990,bid_observation_2010}. 
Even more advanced theoretical proposals exist \cite{wang_edge_2013-1}, which can explain the experimental findings of these states. How to interpret localizations in the case of such a complicated edge structure remains an open question. 
The weak visibility of the $\nu$=2/3 state could be due to this complicated edge structure and suggests a smaller energy gap than observed for the $\nu=1/3$ state.\\
As mentioned before, resonances with bending slopes in the $B$-field$-V_\mathrm{QPC}$ plane are expected for single-electron resonances \cite{ilani_microscopic_2004,jain_quantum_1988}. The detailed behavior of the slope depends on the disorder potential intersecting the Fermi energy. 
This suggests that the resonances in the FQH regime of Fig. \ref{Schmetterling}.a (marked by white dashed line) could be interpreted as single-electron effects. In Ref. \onlinecite{simmons_resistance_1991}, a model for a disorder potential maximum in a constriction is proposed, leading to magnetically bound states which could qualitatively reproduce the bending of the resonances. In this situation, disorder dominates over the smaller FQH gaps and the formation of wide compressible and incompressible regions in the constriction is no longer possible (Fig. \ref{Schema}.D). Thus, the slope in the $B$-field - $V_\mathrm{QPC}$ plane depends on the influence of the QPC voltage on the enclosed area, which depends on the shape of the disorder potential maximum.\\
\subsubsection{B-field and voltage periodicities}
Within this framework, we may now investigate the periodicities of the resonances in Fig. \ref{Faecher}.c-\ref{Faecher}.f. For a Coulomb dominated quantum dot, a distinct behavior of the periodicities $\Delta B(\nu_\mathrm{QPC})$ and $\Delta V_\mathrm{QPC}(\nu_\mathrm{QPC})$ is expected. These periodicities depend on the filling factor of the incompressible region, in which the Coulomb dominated region is formed, in our case this is $\nu_\mathrm{QPC}$. 
From theoretical models \cite{rosenow_influence_2007, halperin_theory_2011} for Coulomb dominated Fabry-P\'erot interferometers it is expected that $\Delta B(\nu_\mathrm{QPC}=1)\approx2 \Delta B(\nu_\mathrm{QPC}=2)\approx\Delta B(\nu_\mathrm{QPC}=1/3)$ and $\Delta V_\mathrm{QPC}(\nu_\mathrm{QPC}=1)\approx\Delta V_\mathrm{QPC}(\nu_\mathrm{QPC}=2)\approx3 \Delta V_\mathrm{QPC}(\nu_\mathrm{QPC}=1/3)$ \footnote{For the voltage periodicity at $\nu_\mathrm{QPC}$=1/3, the gating effect of the background electrons has to be taken into account, as described in Ref. \onlinecite{mcclure_fabry-perot_2012}. In our case however, this only gives a negligible correction from $\Delta V_\mathrm{QPC}(\nu_\mathrm{QPC}=1)\approx3 \Delta V_\mathrm{QPC}(\nu_\mathrm{QPC}=1/3)$.}, which has been observed in lithographically defined quantum dots\cite{mcclure_fabry-perot_2012}. 
In the IQH regime, our periodicities for $\nu_\mathrm{QPC}=2$ ($\Delta B\approx 30~\mathrm{mT},~\Delta V_\mathrm{QPC}\approx 62~mV$) and $\nu_\mathrm{QPC}=1$ ($\Delta B\approx 55~\mathrm{mT},~\Delta V_\mathrm{QPC}\approx 60~mV$) are in good agreement with these predictions. 
Periodicities for $\nu_\mathrm{QPC}=1/3$ ($\Delta B\approx 73~\mathrm{mT},~\Delta V_\mathrm{QPC}\approx 24~mV$) are at least compatible with a Coulomb dominated localization of fractional $e/3$ charges. The area which can be extracted from these periodicities ($A\approx0.075~\mu$m$^2$) is compatible with a localization in the channel of the QPC.
However, it should be noted that the geometry of the compressible region of enhanced or reduced density within the constriction might change as the $B$-field is varied, because it is not lithographically defined but might change self-consistently.
A different behavior is observed in the low-$n$/low-$B$-field end of conductance plateaus ('D' in Fig. \ref{Schema}), where single-electron physics is expected to dominate.
In the measurement of Fig. \ref{Faecher}.c (regions encircled by white dashed line), periodicities for $\nu_\mathrm{QPC}=1$ ($\Delta B\approx 200~\mathrm{mT},~\Delta V_\mathrm{QPC}\approx 53~mV$) and $\nu_\mathrm{QPC}=1/3$ ($\Delta B\approx 360~\mathrm{mT},~\Delta V_\mathrm{QPC}\approx 48~mV$) are incompatible with a Coulomb dominated mechanism and indicate single-electron behavior. Similar enhancements of $\Delta B$ for $\nu=1/3$ have been interpreted as magnetically bound states in earlier experiments \cite{simmons_resistance_1989}. However in this interpretation, finite temperature effects or an interplay with Coulomb blockade mechanisms might have to be taken into account \cite{lee_comment_1990}.\\
\subsubsection{Summary}
To summarize, the most important findings of this section are: periodic conductance oscillations with a slope, parallel to either of the neighboring conductance plateaus were observed. They were interpreted to originate from the Coulomb dominated charging of compressible region of enhanced or reduced density, formed in a constant filling factor background. This filling factor determines the slope. $B$-field and gate voltage periodicities agree with expectations for a Coulomb dominated Fabry-P\'erot interferometer. 
At low densities and in weak magnetic fields, disorder prevents the formation of compressible and incompressible regions. Here, resonances are interpreted as single-electron effects, where electronic states are dominated by confinement and encircle a local potential maximum and enclose a certain number of flux quanta.
In the fractional quantum Hall regime where energy gaps are smaller than in the integer quantum Hall regime, an influence of both mechanisms can be seen. At the plateau boundaries of the $\nu_\mathrm{QPC}$=1/3 state, conductance oscillations, compatible with Coulomb dominated charging of fractionally charged quasiparticles, are observed. For 1/3 $<$ $\nu_\mathrm{QPC}$ $<$ 1, modulations of the conductance with an intermediate slope (in-between slopes of the $\nu_\mathrm{QPC}$=1 and $\nu_\mathrm{QPC}$=1/3 plateaus) are observed. These slopes move with the local filling factor of the QPC, i.e. correspond to a certain number of flux quanta per electron. This indicates the importance of single electron interference, where resonances are expected to emanate from the $B$ = 0, $n$ = 0 origin of the Landau fan \cite{sohrmann_compressibility_2007}.

\subsection{Spatial dependence of QPC resonances}
\label{AsymmetrieChap}
\begin{figure*}
\begin{center}
\includegraphics[width=16cm]{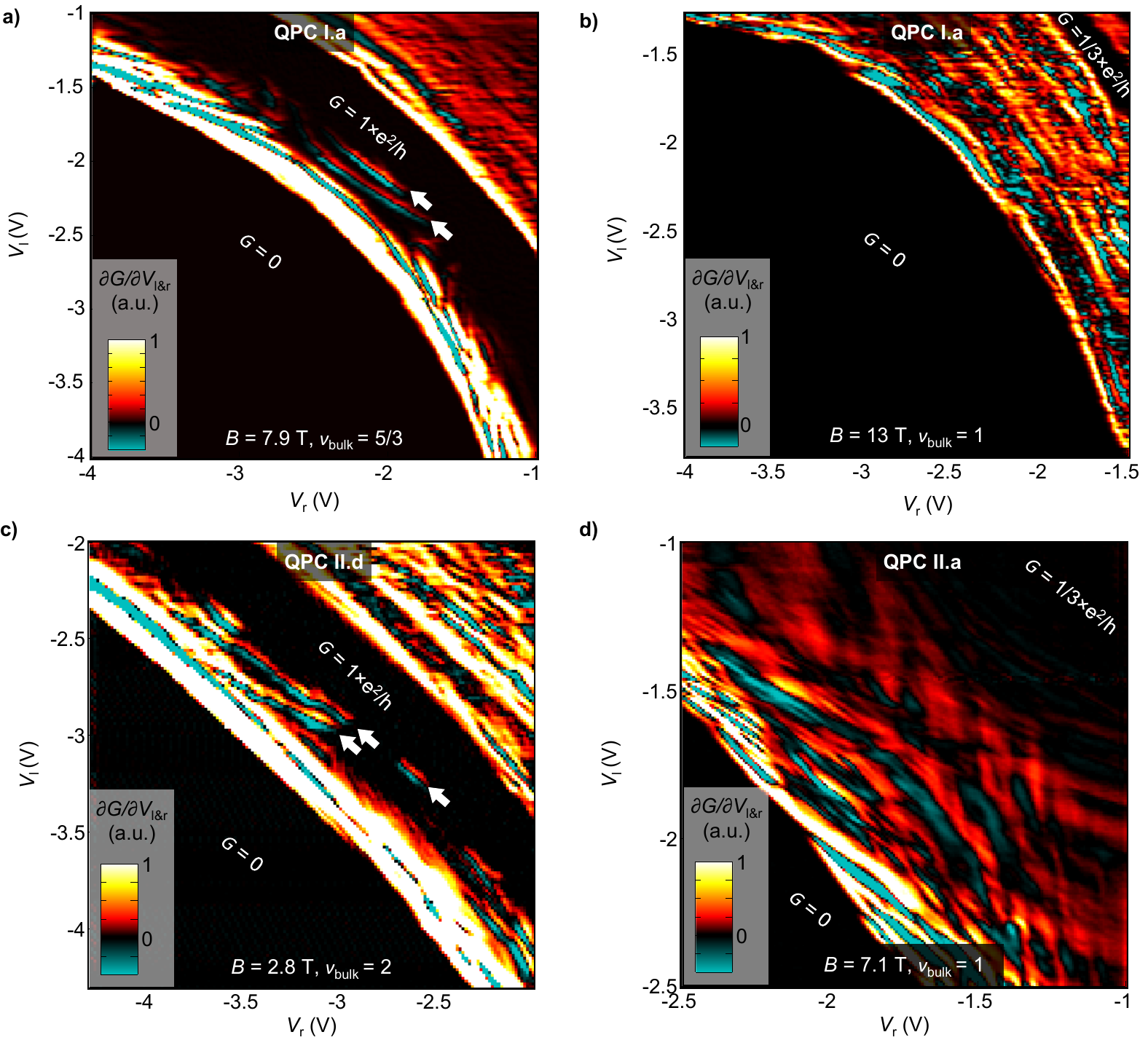}
\end{center}
\caption{(Color online) \textbf{a-d}: Transconductance (numerical derivative in diagonal direction) of QPCs on a high-density sample (QPC I.a, \textbf{a},\textbf{b}) and a low-density sample (QPC II.d, \textbf{c}, QPC II.a, \textbf{d}) as the voltages of left and right QPC gates $V_\mathrm{l}$ and $V_\mathrm{r}$ are varied. White arrows mark resonances that are believed to be due to single-electron interference. These resonances move with a more complicated dependence as the asymmetry is varied, in contrast to many-electron resonances which bend parallel to the conductance plateaus.}
\label{Asymmetrie}
\end{figure*}
 By applying different voltages to the two different QPC gates, it is possible to laterally shift the QPC channel in the lithographically defined constriction (this technique was for example used in references \onlinecite{sarkozy_zero-bias_2009,rossler_transport_2011,komijani_origins_2013}). For QPCs similar to QPC I.a, this shift was found to be of the order of the lithographic QPC width \cite{rossler_transport_2011,schnez_imaging_2011}. 
Fig. \ref{Asymmetrie} shows the numerical derivative of $G$ in diagonal direction (transconductance $\partial G/\partial V_\mathrm{l\&r}$), as the voltages $V_\mathrm{l}$ and $V_\mathrm{r}$ of the left and right QPC gate are varied. In these measurements, the 2DEG far away from the QPC (bulk) is tuned to a fixed filling factor $\nu_\mathrm{bulk}$ with $R_\mathrm{xx}\approx 0$. 
Regions of constant conductance and pinch-off show up as black areas, bright regions of increasing conductance bend around the pinch-off region. 
Fig. \ref{Asymmetrie}.a and \ref{Asymmetrie}.c show the asymmetry-dependence of resonances (the diagonal of Fig. \ref{Asymmetrie}.a is a cut across the resonances of Fig. \ref{Schmetterling}.a indicated by the dashed line) in the low-density low-$B$-field end of the $\nu_\mathrm{QPC}=1$ plateau for three different QPCs (Fig. \ref{Asymmetrie}.a, \ref{Asymmetrie}.b: QPC I.a, Fig. \ref{Asymmetrie}.c: QPC II.d, \ref{Asymmetrie}.d: QPC II.a) on 2DEGs of different density. Resonances believed to originate from single-electron effects (indicated by white arrows), are observed at the low-density end of the $G=1\times e^2/h$ conductance plateau.
The resonances show up as two or three parallel lines with a varying slope clearly different from the conductance plateaus' slope and sit deep in the $G=1\times e^2/h$ conductance plateau. Such resonances, occurring mainly in symmetric configurations, have been observed in most of the QPCs in study.
Additional modulations of the conductance can be observed between the conductance plateaus. These many-electron resonances bend roughly in the same way as the pinch-off line but vary in intensity, as the asymmetry is varied. \\
Fig. \ref{Asymmetrie}.b and \ref{Asymmetrie}.d show the asymmetry behavior in strong magnetic fields. 
For a bulk filling factor $\nu_\mathrm{bulk}=1$, conductance plateaus in the QPC at $G=1/3\times e^2/h$ are observed. 
In Fig. \ref{Asymmetrie}.b, mainly resonances bending with the pinch-off line are observed. 
In contrast, in Fig. \ref{Asymmetrie}.d, non-regular resonances without any preferred slope are observed.\\
With our model (Fig. \ref{Schema}) we can now try to distinguish the asymmetry-behavior of the two different types of resonances: on the one hand, the confinement dominated resonances (Fig. \ref{Schema}.D) for the situation where compressible and incompressible regions are absent and the system is described by single-electron physics, on the other hand the many-electron resonances (Fig. \ref{Schema}.B, \ref{Schema}.C) where a compressible region of enhanced or reduced density, situated in an incompressible region, is charged. \\
Confinement-dominated single-electron resonances are expected to occur as a result of a localized state at a certain position in the channel, to which both edges couple. As the asymmetry and thus the background potential is varied, single-particle energy levels are shifted in energy, which changes the position in gate voltages of the resonance relative to pinch-off. Thus, single-electron resonances are expected to possess a dependence on gate voltage which is not parallel to the respective conductance plateau as the asymmetry is varied. They should disappear, as soon as the coupling to one of the edges is lost. Here, the gate voltage dependence is influenced by the details of the confinement and disorder potential. 
The proposed gate voltage dependence of the single-electron resonances (which causes a bending not necessarily parallel to the pinch-off line) and the disappearance of the resonances with increasing asymmetry are indeed observed (Fig.  \ref{Asymmetrie}.a, \ref{Asymmetrie}.b, white arrows). 
A similar behavior might be expected from an Aharonov-Bohm mechanism, where non-adiabaticity of the QPC saddle-point potential leads to enhanced tunneling between the edge channels at the entrance and exit of the constriction and thus defines a QPC-voltage dependent area \cite{van_loosdrecht_aharonov-bohm_1988}.\\

In the ideal model of many-electron resonances, the charge of the compressible island of reduced or enhanced density is quantized and changes when the total density in the constriction is varied (i.e. when moving perpendicular to pinch-off in the $V_\mathrm{l}-V_\mathrm{r}$ plane). When the asymmetry is varied parallel to pinch-off, we expect to change mainly the width of the incompressible regions separating the compressible island of reduced or enhanced density from the edge. Thereby the resonance amplitude which highly depends on the width of the incompressible region \cite{baer_cyclic_2013} is changed. At the same time, the occupation of the compressible region of reduced or enhanced density is expected to be approximately constant, as long as the picture of compressible and incompressible regions does not break down. 
In this scenario, resonances are thus expected to run parallel to the conductance plateau edges, as observed in the measurements (Figs. \ref{Asymmetrie}.a-c). \\
Because the conductance varies strongly in-between the plateaus, resonances cannot be attributed to individual localizations as it was possible for example in Fig. \ref{Faecher}.c. 
Thus, in a yet different scenario, conductance oscillations could also originate from single-electron effects, where we only probe localizations that couple to both edges for a given voltage asymmetry. At this asymmetry, they possess a local gate voltage dependence, shifting them parallel to the conductance plateaus. 
The overall behavior of the resonances could result from averaging the contributions of many single-electron resonances.\\
Summarizing, we may state that the bending resonances of Fig. \ref{Asymmetrie}.a and \ref{Asymmetrie}.c (marked by white arrows) are compatible with a confinement dominated single-electron effect, whereas resonances parallel to the conductance plateaus (Figs. \ref{Asymmetrie}.a-.c) are compatible with a many-electron effect. However, other mechanisms leading to similar observations cannot be excluded.
The fact that in Fig. \ref{Asymmetrie}.d no resonances bending with the conductance plateaus are observed may indicate that in Fig. \ref{Asymmetrie}.d transport  is dominated by single-electron physics, while many-electron effects dominate in Fig. \ref{Asymmetrie}.b, where the applied magnetic field is much stronger and the disorder potential is smaller due to a higher mobility 2DEG.


\subsection{Fragile fractional quantum Hall states in QPCs}
\label{Fragile}
\begin{figure}
\begin{center}
\includegraphics[width=8cm]{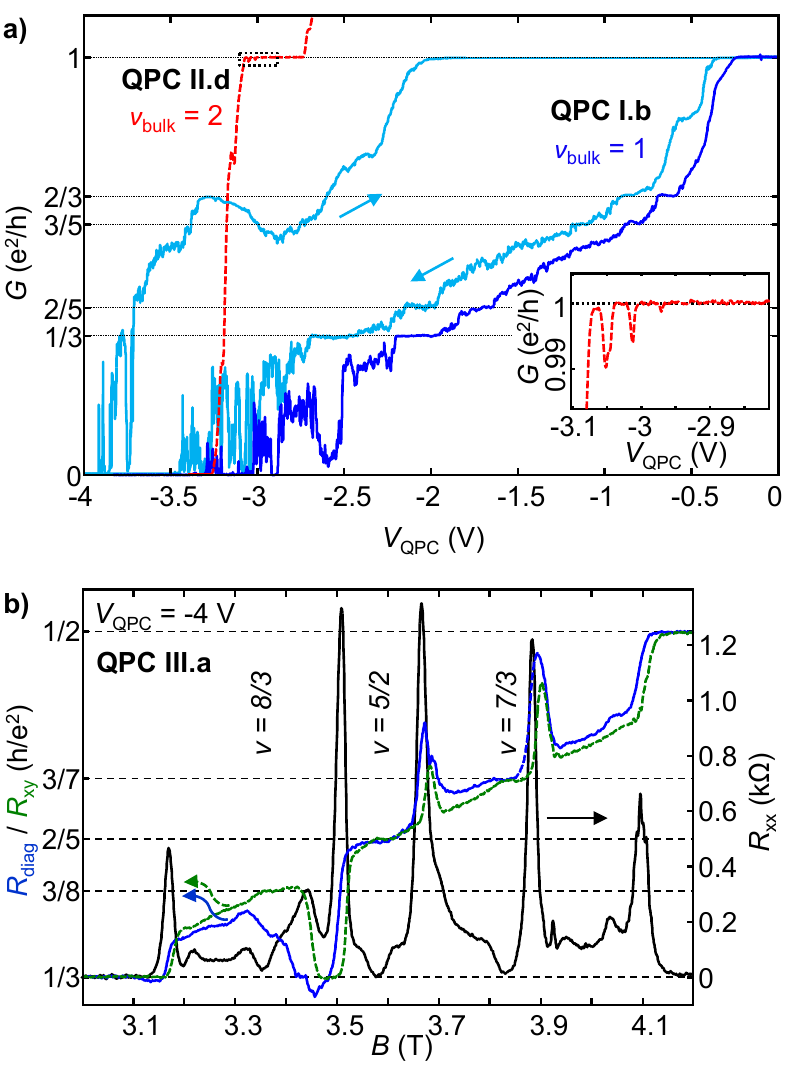}
\end{center}
\caption{(Color online) \textbf{a}: In strong magnetic fields ($B$ = 13 T), the transmission of QPC I.b close to pinch-off is strongly fluctuating (light / dark blue (gray)). The dashed (red) curve depicts a situation in which a transmitted edge state is weakly backscattered in QPC II.d (see inset). \textbf{b}: Transmission of QPC III.a for 2 $\le$ $\nu_\mathrm{bulk}$ $\le$ 3. In the bulk, $\nu_\mathrm{bulk}=$ 7/3, 8/3 and 5/2 are fully quantized with a strong minimum in $R_{\mathrm{xx}}$ (solid black line) and a plateau in $R_{\mathrm{xy}}$ (dashed (green) line). In addition, pronounced reentrant integer quantum Hall (RIQH) states are observed. The diagonal resistance across the QPC, $R_\mathrm{diag}$ (solid blue (gray) line), shows a plateau at $\nu=5/2$, indicating nearly perfect transmission through the QPC. The density within the constriction is very similar to the bulk density.}
\label{Cold}
\end{figure}

Fig. \ref{Cold}.a shows the transmission of QPC I.b (light / dark blue) and QPC II.d (red) as a function of applied QPC voltage. 
The conductance of QPC II.d as a function of $V_\mathrm{QPC}$ (red) shows conductance oscillations on the low-density side of the $\nu_\mathrm{QPC}=1$ plateau. These are those resonances of Fig.\ref{Asymmetrie}.c, which were interpreted as single-electron effects.\\
QPC I.b exhibits conductance plateaus at 2/3, 3/5, 2/5 and $1/3\times e^2/h$ in strong magnetic fields ($B = 13~\mathrm{T}$, $\nu_\mathrm{bulk}=1$). Close to pinch-off, the conductance strongly fluctuates. Unfortunately the observation of $\nu_\mathrm{QPC}=2/3,~3/5,~2/5$ and 1/3 does not allow to draw conclusions about the edge reconstruction of the $\nu_\mathrm{bulk}=1$ edge state. Over the whole QPC voltage range, not only the transmission, but also the channel density and the shape of the QPC confinement potential strongly vary \cite{rossler_transport_2011}.
The measurement in dark blue shows the first $V_\mathrm{QPC}$ sweep after the cool-down. When closing the channel for a second time (light blue (gray)), fractional filling factors are still visible, but a more negative gate voltage has to be applied to pinch off the channel. As the QPC is subsequently opened again, a pronounced hysteresis is visible and the more fragile conductance plateaus at $G=2/5\times e^2/h$, $G=3/5\times e^2/h$, $G=1/3\times e^2/h$ and $G=2/3\times e^2/h$ disappear. 
This behavior can be understood considering the time- and voltage-dependent density in the X-electron screening layers. After the screening layers have been depleted, the density only relaxes with long time constants. The electron density of the 2DEG is inversely proportional to the charge carrier density in the X-electron bands due to capacitive coupling. Thus, depleted screening layers lead to a increased 2DEG electron density at the same QPC voltage, explaining why the QPC conductance is higher for opening the QPC than for closing it. 

Fig. \ref{Cold}.a, demonstrates that many different fractional filling factors $\nu_\mathrm{QPC}$ can be transmitted by applying an appropriate QPC voltage and keeping the magnetic field fixed. 
However, relaxation of the barely mobile X-band screening layer electrons makes the observation of the most fragile fractional quantum Hall states difficult. 
To overcome this limitation, the fact that the X-band screening layers become mobile for temperatures above approximately 1 K can be used \cite{rossler_gating_2010,radu_quasi-particle_2008}. 
By applying top-gate voltages at higher temperatures, the screening layer density can relax in a steady state and density fluctuations in the constriction are avoided. By this relaxation, additional screening is provided, which is believed to result in a much steeper QPC confinement potential.
The density relaxation is extremely slow at dilution refrigerator temperatures. At $T$ $\approx$ 1 K, the density already saturates within minutes. At $T$ $\approx$ 4 K, relaxation to lower screening electron density is nearly instantaneous. To allow a full relaxation of the screening layers, the system is kept at $T$ $\approx$ 4 K for several hours. If in contrast the same gate voltage is applied at mK temperatures, the channel density is strongly reduced due to the vicinity of negatively charged screening electrons. Simultaneously to the slow depletion of the screening layers, the electron density in the channel rises. However, this process leads to strong fluctuations in the conductance which destroy the quantization of fragile fractional quantum Hall states.\\
Fig. \ref{Cold}.b shows the diagonal resistance of the 1.2 $\mu$m wide QPC III.a for 2 $\le$ $\nu_\mathrm{bulk}$ $\le$ 3. Here, -4 V have been applied to the QPC gates at $T$ $\approx$ 4 K. 
The electron gas below the metallic top-gates is depleted at approx. -3.2 V. At a base temperature of 9 mK (electronic temperature $\approx$ 12 mK), the filling factors 7/3, 8/3 and 5/2 are fully quantized in the bulk, with a strong minimum in $R_{\mathrm{xx}}$ and a plateau in $R_{\mathrm{xy}}$. 
In addition, pronounced reentrant integer quantum Hall states are observed. 
The density in the constriction is nearly identical to the bulk density, as seen from the overlap of different filling factors. 
At a magnetic field of approx. 3.6 T, the plateau in $R_\mathrm{diag}$ shows, that the $\nu=5/2$ state is nearly perfectly transmitted through the QPC, without significant backscattering. Here, the applied QPC voltage of -4 V has been kept fixed while cooling down to the base temperature.
The deviation of $R_\mathrm{diag}$ and $R_\mathrm{xy}$ at $B$ $\approx$ 3.4 T originates from a small longitudinal component in $R_\mathrm{diag}$ due to an asymmetry of the sample geometry. 
The optimized growth and gating procedure allow the definition of a QPC without decreasing the density in the constriction and without destroying the quantization of the $\nu$=5/2 state, which is otherwise not possible. 
Interference experiments at $\nu$=5/2 \cite{stern_proposed_2006,bonderson_detecting_2006,bonderson_probing_2006,ilan_signatures_2011,ilan_coulomb_2008,willett_measurement_2009,willett_magnetic-field-tuned_2013} require a filling factor $\nu$=5/2 in the center of the employed QD, while edge states are only partially transmitted. Here, the diameter of the QD is constrained to a few $\mu$m (due to the finite quasiparticle coherence length \cite{bishara_interferometric_2009}), thus making the conservation of the bulk density and $\nu$=5/2 quantization on a $\mu$m length-scale crucial. 
The steep confinement potential of QPCIII.a leads to a decreased width of the compressible regions in the QPC and a wider separating incompressible region, thus reducing backscattering across. The anticipated complex edge structure of the $\nu$ = 5/2 state (which was experimentally found to occur only in QPCs of rather large width \cite{miller_fractional_2007}) might facilitate its formation in a steeper confinement potential. 
Furthermore, the additional screening of the disorder in the constriction via X-band electrons reduces the amplitude of the disorder potential fluctuations. Hence, the influence of conductance oscillations as discussed in section \ref{main} is expected to be reduced.
The main drawback of the utilized gating method is the low tuneability of gate voltages at mK temperatures. Here, the gate voltages have to remain in very small range around the voltage that has been applied at $T$=4 K. Otherwise, slow relaxation processes of the X-band screening layers destroy the quantization of the $\nu$=5/2 state. Growth methods which utilize conventional DX-doping and a reduced Al molar fraction might help to overcome this problem, while still providing a good quantization of $\nu$=5/2 \cite{reichl_increasing_2014,pan_impact_2011,gamez_5/2_2013}.
\\
Having demonstrated that we can confine a fully gapped $\nu$=5/2 state to a QPC, we are at a good starting point for conducting tunneling and interference experiments with the fragile fractional quantum Hall states at $\nu$=7/3 and 5/2.

\section{Conclusion}
In conclusion, we have investigated the interplay of electronic transport and localization in quantum point contacts of different geometries and based on 2DEGs utilizing different growth techniques. 
In these systems, various integer and fractional quantum Hall states were observed. Using a QPC with a top-gate, we were able to investigate conductance resonances in greater detail. 
In this sample, edge states are separated by a wide incompressible region thus leading to a significant influence of localizations due to disorder potential fluctuations.
Regions of perfect QPC transmission are surrounded by periodic conductance oscillations with an identical slope in the $V_\mathrm{QPC}-B$-field plane. 
Within a many-electron picture, the resonances on the high (low) density end of the plateau can be interpreted as regions of enhanced or reduced density formed within incompressible regions between the counter-propagating edge states. 
As the charge of these regions is conserved, changing the density or magnetic field leads to periodic conductance oscillations, whenever an electron is added or removed. 
$B$-field and $V_\mathrm{QPC}$-periodicities agree with expectations for a Coulomb dominated quantum dot in strong magnetic fields and are determined by the filling factor background in which the compressible region of enhanced or reduced density is formed.
At low densities and in weaker magnetic fields, resonances within the conductance plateau occur. 
In this regime, disorder broadening becomes comparable to the Landau level separation, thus compressible regions of reduced and enhanced density, situated in different Landau levels modulate transport at the same time. Here, the many-electron picture is not valid anymore and resonances with a dependence in the $B-V_\mathrm{QPC}$ plane, not necessarily equal to the conductance plateaus' dependence, are observed. These resonances are interpreted as confinement dominated single-electron interference effects.
In the fractional quantum Hall regime, the behavior of the system seems to be influenced by both, single- and many-electron physics. 
Due to the much smaller gaps of the FQH states, disorder becomes more important. 
Close to perfect transmission, resonances similar to those associated with compressible regions of reduced or enhanced density in a many-electron picture can be observed. Periodicities at $\nu_\mathrm{QPC}=1/3$ are compatible with the localization of fractionally charged quasiparticles in a Coulomb dominated quantum dot.
However, for intermediate transmissions, weak resonances with a slope different from the slopes of the neighboring conductance plateaus are observed, indicating the importance of single-electron physics where the formation of compressible and incompressible regions breaks down.
Single-electron resonances have been studied as a function of the position of the conducting channel in the constriction. In contrast to many-electron resonances, single-electron resonances are expected to possess slopes in the gate-voltage plane, not necessarily parallel to the conductance plateaus. Here, the slope depends on the details of the disorder potential.
Using optimized growth techniques and gating procedures, we are able to form QPC constrictions with extremely weak backscattering and a density equal to the bulk density. This allows us to observe the $\nu=5/2$ state in the QPC with a fully developed plateau. The bulk properties, like the reentrant integer quantum Hall states, are fully conserved in the QPC, making this system promising for future tunneling and interference experiments at $\nu$=5/2.

\section{Acknowledgments}
We gratefully acknowledge discussions with Bernd Rosenow, Yigal Meir, Yuval Gefen, Ferdinand Kuemmeth and Hiske Overweg. We acknowledge the support of the ETH FIRST laboratory and financial support of the
Swiss Science Foundation (Schweizerischer Nationalfonds, NCCR 'Quantum Science and Technology').

\appendix*
\section{Energy gap of the $\nu_\mathrm{QPC}$ = 1/3 state}

\begin{figure}
\begin{center}
\includegraphics[width=8cm]{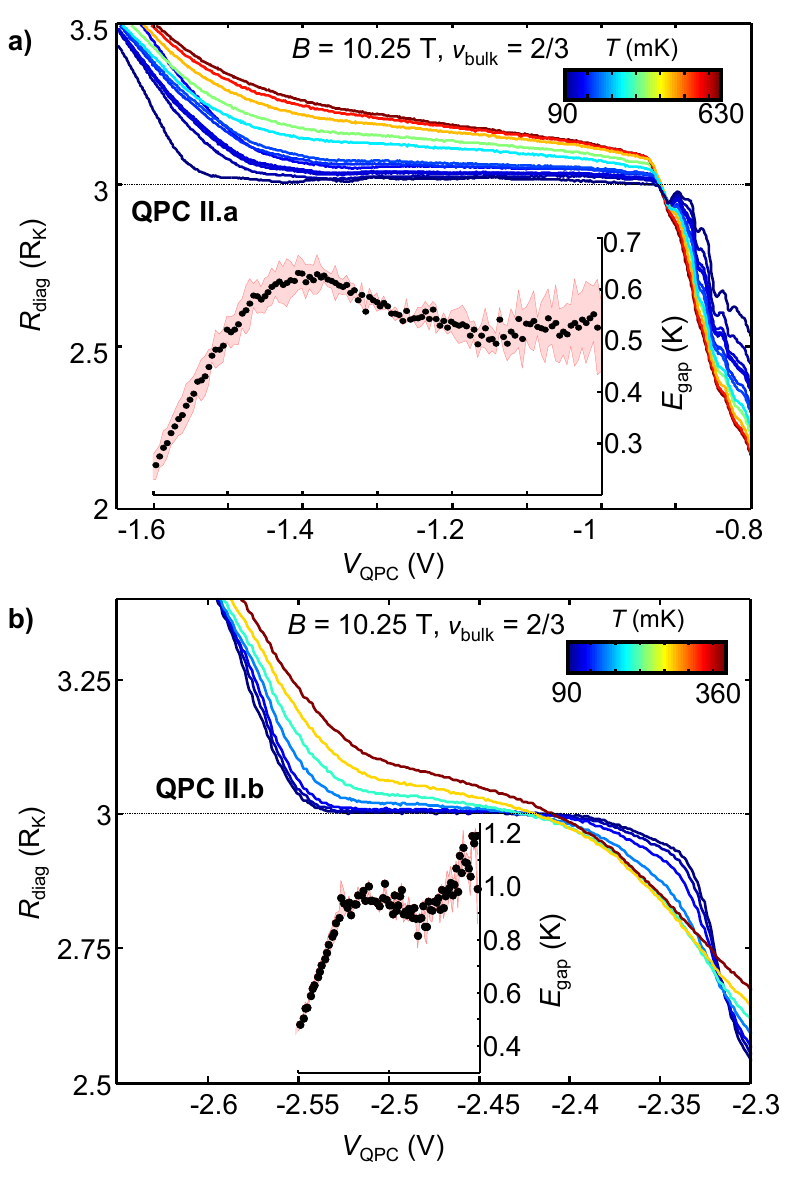}
\end{center}
\caption{(Color online) \textbf{a:} Diagonal resistances for different temperatures at $\nu_\mathrm{bulk}=2/3$ for QPC II.a (here $V_\mathrm{CTG}=-0.47~\mathrm{V}$) and QPC II.b (\textbf{b}). The insets show the energy gap as a function of the QPC gate voltage, which has been extracted from the activated behavior of $\Delta R_\mathrm{diag}$. The shaded area depicts an estimate of the fit error for $E_\mathrm{gap}$.}
\label{Activation}
\end{figure}

Activation measurements have been performed on the $\nu_\mathrm{QPC}$ = 1/3 states in the two QPCs QPCII.a and QPCII.b. The measured diagonal resistances $R_\mathrm{diag}$ of both QPCs at a magnetic field of 10.25 T are shown in Figs. \ref{Activation}.a and \ref{Activation}.b. 
Here, a two-terminal AC voltage modulation of $V_\mathrm{AC}$ = 40 $\mu$V, corresponding to an AC current $I_\mathrm{AC}$ of approximately 0.5 nA has been applied.
The plateau at $R_\mathrm{diag}=3\times R_\mathrm{K}$ ($R_\mathrm{K}=h/e^2$), corresponding to $\nu_\mathrm{QPC}=1/3$, is much wider for QPC II.a (Fig. \ref{Activation}.a).
 Temperature-dependent measurements reveal an activated behavior $\Delta R_\mathrm{diag}\varpropto e^{-\frac{\Delta_\mathrm{diag}}{k_BT}}$ of the deviation of the diagonal resistance from its plateau value, $\Delta R_\mathrm{diag}$. 
The energy gap values, extracted at different QPC voltages, are shown as insets in Fig.\ref{Activation}.a and \ref{Activation}.b. Extracted energy gaps ($E_\mathrm{gap}=2\Delta_\mathrm{diag}$) correspond to thermal energies between 0.6 K and 1.0 K for these two QPCs, compared to an energy gap of 3.2 K for the bulk $\nu$ = 2/3 state at the same magnetic field.
Thus, $R_\mathrm{xx}\approx 0$ has been maintained over the whole temperature range in the bulk ($\nu_\mathrm{bulk}$=2/3), meaning that we probe only the temperature-dependence of the QPC. 
In contrast to activation measurements of $R_\mathrm{xy}$ in the bulk\cite{wei_temperature_1985}, deviations from the quantized resistance value do not occur symmetrically around the center of the plateau. 
This effect, which is believed to be due to electron-electron interactions \cite{siddiki_interaction-mediated_2009}, is much more pronounced in the QPC with the CTG. 
The similar size of these energy gaps suggests that the different widths and shapes of the $\nu_\mathrm{QPC}$=1/3 plateau (as $V_\mathrm{QPC}$ is varied) mainly stem from different shapes of the confinement potential. 

\bibliographystyle{apsrev}
\bibliography{Bibliothek}

\begin{thebibliography}{90}
\expandafter\ifx\csname natexlab\endcsname\relax\def\natexlab#1{#1}\fi
\expandafter\ifx\csname bibnamefont\endcsname\relax
  \def\bibnamefont#1{#1}\fi
\expandafter\ifx\csname bibfnamefont\endcsname\relax
  \def\bibfnamefont#1{#1}\fi
\expandafter\ifx\csname citenamefont\endcsname\relax
  \def\citenamefont#1{#1}\fi
\expandafter\ifx\csname url\endcsname\relax
  \def\url#1{\texttt{#1}}\fi
\expandafter\ifx\csname urlprefix\endcsname\relax\def\urlprefix{URL }\fi
\providecommand{\bibinfo}[2]{#2}
\providecommand{\eprint}[2][]{\url{#2}}

\bibitem[{\citenamefont{Komiyama}(1998)}]{komiyama_edge_1998}
\bibinfo{author}{\bibfnamefont{S.}~\bibnamefont{Komiyama}}, in
  \emph{\bibinfo{booktitle}{Mesoscopic Physics and Electronics}}, edited by
  \bibinfo{editor}{\bibfnamefont{P.~T.} \bibnamefont{Ando}},
  \bibinfo{editor}{\bibfnamefont{P.~Y.} \bibnamefont{Arakawa}},
  \bibinfo{editor}{\bibfnamefont{P.~K.} \bibnamefont{Furuya}},
  \bibinfo{editor}{\bibfnamefont{P.~S.} \bibnamefont{Komiyama}},
  \bibnamefont{and} \bibinfo{editor}{\bibfnamefont{P.~H.}
  \bibnamefont{Nakashima}} (\bibinfo{publisher}{Springer Berlin Heidelberg},
  \bibinfo{year}{1998}), {NanoScience} and Technology, pp.
  \bibinfo{pages}{120--131}, ISBN \bibinfo{isbn}{978-3-642-71978-3,
  978-3-642-71976-9},
  \urlprefix\url{http://link.springer.com/chapter/10.1007/978-3-642-71976-9_16}.

\bibitem[{\citenamefont{Tessmer et~al.}(1998)\citenamefont{Tessmer,
  Glicofridis, Ashoori, Levitov, and Melloch}}]{tessmer_subsurface_1998}
\bibinfo{author}{\bibfnamefont{S.~H.} \bibnamefont{Tessmer}},
  \bibinfo{author}{\bibfnamefont{P.~I.} \bibnamefont{Glicofridis}},
  \bibinfo{author}{\bibfnamefont{R.~C.} \bibnamefont{Ashoori}},
  \bibinfo{author}{\bibfnamefont{L.~S.} \bibnamefont{Levitov}},
  \bibnamefont{and} \bibinfo{author}{\bibfnamefont{M.~R.}
  \bibnamefont{Melloch}}, \bibinfo{journal}{Nature}
  \textbf{\bibinfo{volume}{392}}, \bibinfo{pages}{51} (\bibinfo{year}{1998}),
  ISSN \bibinfo{issn}{0028-0836},
  \urlprefix\url{http://www.nature.com/nature/journal/v392/n6671/full/392051a0.html}.

\bibitem[{\citenamefont{Ilani et~al.}(2004)\citenamefont{Ilani, Martin,
  Teitelbaum, Smet, Mahalu, Umansky, and Yacoby}}]{ilani_microscopic_2004}
\bibinfo{author}{\bibfnamefont{S.}~\bibnamefont{Ilani}},
  \bibinfo{author}{\bibfnamefont{J.}~\bibnamefont{Martin}},
  \bibinfo{author}{\bibfnamefont{E.}~\bibnamefont{Teitelbaum}},
  \bibinfo{author}{\bibfnamefont{J.~H.} \bibnamefont{Smet}},
  \bibinfo{author}{\bibfnamefont{D.}~\bibnamefont{Mahalu}},
  \bibinfo{author}{\bibfnamefont{V.}~\bibnamefont{Umansky}}, \bibnamefont{and}
  \bibinfo{author}{\bibfnamefont{A.}~\bibnamefont{Yacoby}},
  \bibinfo{journal}{Nature} \textbf{\bibinfo{volume}{427}},
  \bibinfo{pages}{328} (\bibinfo{year}{2004}), ISSN \bibinfo{issn}{0028-0836},
  \urlprefix\url{http://www.nature.com/nature/journal/v427/n6972/abs/nature02230.html}.

\bibitem[{\citenamefont{Martin et~al.}(2004)\citenamefont{Martin, Ilani,
  Verdene, Smet, Umansky, Mahalu, Schuh, Abstreiter, and
  Yacoby}}]{martin_localization_2004}
\bibinfo{author}{\bibfnamefont{J.}~\bibnamefont{Martin}},
  \bibinfo{author}{\bibfnamefont{S.}~\bibnamefont{Ilani}},
  \bibinfo{author}{\bibfnamefont{B.}~\bibnamefont{Verdene}},
  \bibinfo{author}{\bibfnamefont{J.}~\bibnamefont{Smet}},
  \bibinfo{author}{\bibfnamefont{V.}~\bibnamefont{Umansky}},
  \bibinfo{author}{\bibfnamefont{D.}~\bibnamefont{Mahalu}},
  \bibinfo{author}{\bibfnamefont{D.}~\bibnamefont{Schuh}},
  \bibinfo{author}{\bibfnamefont{G.}~\bibnamefont{Abstreiter}},
  \bibnamefont{and} \bibinfo{author}{\bibfnamefont{A.}~\bibnamefont{Yacoby}},
  \bibinfo{journal}{Science} \textbf{\bibinfo{volume}{305}},
  \bibinfo{pages}{980} (\bibinfo{year}{2004}), ISSN \bibinfo{issn}{0036-8075,
  1095-9203}, \bibinfo{note}{{PMID:} 15310895},
  \urlprefix\url{http://www.sciencemag.org/content/305/5686/980}.

\bibitem[{\citenamefont{Steele et~al.}(2005)\citenamefont{Steele, Ashoori,
  Pfeiffer, and West}}]{steele_imaging_2005}
\bibinfo{author}{\bibfnamefont{G.~A.} \bibnamefont{Steele}},
  \bibinfo{author}{\bibfnamefont{R.~C.} \bibnamefont{Ashoori}},
  \bibinfo{author}{\bibfnamefont{L.~N.} \bibnamefont{Pfeiffer}},
  \bibnamefont{and} \bibinfo{author}{\bibfnamefont{K.~W.} \bibnamefont{West}},
  \bibinfo{journal}{Physical Review Letters} \textbf{\bibinfo{volume}{95}},
  \bibinfo{pages}{136804} (\bibinfo{year}{2005}),
  \urlprefix\url{http://link.aps.org/doi/10.1103/PhysRevLett.95.136804}.

\bibitem[{\citenamefont{Hashimoto et~al.}(2008)\citenamefont{Hashimoto,
  Sohrmann, Wiebe, Inaoka, Meier, Hirayama, R{\"o}mer, Wiesendanger, and
  Morgenstern}}]{hashimoto_quantum_2008}
\bibinfo{author}{\bibfnamefont{K.}~\bibnamefont{Hashimoto}},
  \bibinfo{author}{\bibfnamefont{C.}~\bibnamefont{Sohrmann}},
  \bibinfo{author}{\bibfnamefont{J.}~\bibnamefont{Wiebe}},
  \bibinfo{author}{\bibfnamefont{T.}~\bibnamefont{Inaoka}},
  \bibinfo{author}{\bibfnamefont{F.}~\bibnamefont{Meier}},
  \bibinfo{author}{\bibfnamefont{Y.}~\bibnamefont{Hirayama}},
  \bibinfo{author}{\bibfnamefont{R.~A.} \bibnamefont{R{\"o}mer}},
  \bibinfo{author}{\bibfnamefont{R.}~\bibnamefont{Wiesendanger}},
  \bibnamefont{and}
  \bibinfo{author}{\bibfnamefont{M.}~\bibnamefont{Morgenstern}},
  \bibinfo{journal}{Physical Review Letters} \textbf{\bibinfo{volume}{101}},
  \bibinfo{pages}{256802} (\bibinfo{year}{2008}),
  \urlprefix\url{http://link.aps.org/doi/10.1103/PhysRevLett.101.256802}.

\bibitem[{\citenamefont{Wei et~al.}(1998)\citenamefont{Wei, Weis, Klitzing, and
  Eberl}}]{wei_edge_1998}
\bibinfo{author}{\bibfnamefont{Y.~Y.} \bibnamefont{Wei}},
  \bibinfo{author}{\bibfnamefont{J.}~\bibnamefont{Weis}},
  \bibinfo{author}{\bibfnamefont{K.~v.} \bibnamefont{Klitzing}},
  \bibnamefont{and} \bibinfo{author}{\bibfnamefont{K.}~\bibnamefont{Eberl}},
  \bibinfo{journal}{Physical Review Letters} \textbf{\bibinfo{volume}{81}},
  \bibinfo{pages}{1674} (\bibinfo{year}{1998}),
  \urlprefix\url{http://link.aps.org/doi/10.1103/PhysRevLett.81.1674}.

\bibitem[{\citenamefont{Chklovskii et~al.}(1992)\citenamefont{Chklovskii,
  Shklovskii, and Glazman}}]{chklovskii_electrostatics_1992}
\bibinfo{author}{\bibfnamefont{D.~B.} \bibnamefont{Chklovskii}},
  \bibinfo{author}{\bibfnamefont{B.~I.} \bibnamefont{Shklovskii}},
  \bibnamefont{and} \bibinfo{author}{\bibfnamefont{L.~I.}
  \bibnamefont{Glazman}}, \bibinfo{journal}{Physical Review B}
  \textbf{\bibinfo{volume}{46}}, \bibinfo{pages}{4026} (\bibinfo{year}{1992}),
  \urlprefix\url{http://link.aps.org/doi/10.1103/PhysRevB.46.4026}.

\bibitem[{\citenamefont{Efros}(1992)}]{efros_homogeneous_1992}
\bibinfo{author}{\bibfnamefont{A.~L.} \bibnamefont{Efros}},
  \bibinfo{journal}{Physical Review B} \textbf{\bibinfo{volume}{45}},
  \bibinfo{pages}{11354} (\bibinfo{year}{1992}),
  \urlprefix\url{http://link.aps.org/doi/10.1103/PhysRevB.45.11354}.

\bibitem[{\citenamefont{Cooper and Chalker}(1993)}]{cooper_coulomb_1993}
\bibinfo{author}{\bibfnamefont{N.~R.} \bibnamefont{Cooper}} \bibnamefont{and}
  \bibinfo{author}{\bibfnamefont{J.~T.} \bibnamefont{Chalker}},
  \bibinfo{journal}{Physical Review B} \textbf{\bibinfo{volume}{48}},
  \bibinfo{pages}{4530} (\bibinfo{year}{1993}),
  \urlprefix\url{http://link.aps.org/doi/10.1103/PhysRevB.48.4530}.

\bibitem[{\citenamefont{Simmons et~al.}(1989)\citenamefont{Simmons, Wei, Engel,
  Tsui, and Shayegan}}]{simmons_resistance_1989}
\bibinfo{author}{\bibfnamefont{J.~A.} \bibnamefont{Simmons}},
  \bibinfo{author}{\bibfnamefont{H.~P.} \bibnamefont{Wei}},
  \bibinfo{author}{\bibfnamefont{L.~W.} \bibnamefont{Engel}},
  \bibinfo{author}{\bibfnamefont{D.~C.} \bibnamefont{Tsui}}, \bibnamefont{and}
  \bibinfo{author}{\bibfnamefont{M.}~\bibnamefont{Shayegan}},
  \bibinfo{journal}{Physical Review Letters} \textbf{\bibinfo{volume}{63}},
  \bibinfo{pages}{1731} (\bibinfo{year}{1989}),
  \urlprefix\url{http://link.aps.org/doi/10.1103/PhysRevLett.63.1731}.

\bibitem[{\citenamefont{Simmons et~al.}(1991)\citenamefont{Simmons, Hwang,
  Tsui, Wei, Engel, and Shayegan}}]{simmons_resistance_1991}
\bibinfo{author}{\bibfnamefont{J.~A.} \bibnamefont{Simmons}},
  \bibinfo{author}{\bibfnamefont{S.~W.} \bibnamefont{Hwang}},
  \bibinfo{author}{\bibfnamefont{D.~C.} \bibnamefont{Tsui}},
  \bibinfo{author}{\bibfnamefont{H.~P.} \bibnamefont{Wei}},
  \bibinfo{author}{\bibfnamefont{L.~W.} \bibnamefont{Engel}}, \bibnamefont{and}
  \bibinfo{author}{\bibfnamefont{M.}~\bibnamefont{Shayegan}},
  \bibinfo{journal}{Physical Review B} \textbf{\bibinfo{volume}{44}},
  \bibinfo{pages}{12933} (\bibinfo{year}{1991}),
  \urlprefix\url{http://link.aps.org/doi/10.1103/PhysRevB.44.12933}.

\bibitem[{\citenamefont{Cobden et~al.}(1999)\citenamefont{Cobden, Barnes, and
  Ford}}]{cobden_fluctuations_1999}
\bibinfo{author}{\bibfnamefont{D.~H.} \bibnamefont{Cobden}},
  \bibinfo{author}{\bibfnamefont{C.~H.~W.} \bibnamefont{Barnes}},
  \bibnamefont{and} \bibinfo{author}{\bibfnamefont{C.~J.~B.}
  \bibnamefont{Ford}}, \bibinfo{journal}{Physical Review Letters}
  \textbf{\bibinfo{volume}{82}}, \bibinfo{pages}{4695} (\bibinfo{year}{1999}),
  \urlprefix\url{http://link.aps.org/doi/10.1103/PhysRevLett.82.4695}.

\bibitem[{\citenamefont{Machida et~al.}(2001)\citenamefont{Machida, Ishizuka,
  Komiyama, Muraki, and Hirayama}}]{machida_resistance_2001}
\bibinfo{author}{\bibfnamefont{T.}~\bibnamefont{Machida}},
  \bibinfo{author}{\bibfnamefont{S.}~\bibnamefont{Ishizuka}},
  \bibinfo{author}{\bibfnamefont{S.}~\bibnamefont{Komiyama}},
  \bibinfo{author}{\bibfnamefont{K.}~\bibnamefont{Muraki}}, \bibnamefont{and}
  \bibinfo{author}{\bibfnamefont{Y.}~\bibnamefont{Hirayama}},
  \bibinfo{journal}{Physical Review B} \textbf{\bibinfo{volume}{63}},
  \bibinfo{pages}{045318} (\bibinfo{year}{2001}),
  \urlprefix\url{http://link.aps.org/doi/10.1103/PhysRevB.63.045318}.

\bibitem[{\citenamefont{Couturaud et~al.}(2009)\citenamefont{Couturaud,
  Bonifacie, Jouault, Mailly, Raymond, and Chaubet}}]{couturaud_local_2009}
\bibinfo{author}{\bibfnamefont{O.}~\bibnamefont{Couturaud}},
  \bibinfo{author}{\bibfnamefont{S.}~\bibnamefont{Bonifacie}},
  \bibinfo{author}{\bibfnamefont{B.}~\bibnamefont{Jouault}},
  \bibinfo{author}{\bibfnamefont{D.}~\bibnamefont{Mailly}},
  \bibinfo{author}{\bibfnamefont{A.}~\bibnamefont{Raymond}}, \bibnamefont{and}
  \bibinfo{author}{\bibfnamefont{C.}~\bibnamefont{Chaubet}},
  \bibinfo{journal}{Physical Review B} \textbf{\bibinfo{volume}{80}},
  \bibinfo{pages}{033304} (\bibinfo{year}{2009}),
  \urlprefix\url{http://link.aps.org/doi/10.1103/PhysRevB.80.033304}.

\bibitem[{\citenamefont{Peled et~al.}(2003)\citenamefont{Peled, Shahar, Chen,
  Diez, Sivco, and Cho}}]{peled_near-perfect_2003}
\bibinfo{author}{\bibfnamefont{E.}~\bibnamefont{Peled}},
  \bibinfo{author}{\bibfnamefont{D.}~\bibnamefont{Shahar}},
  \bibinfo{author}{\bibfnamefont{Y.}~\bibnamefont{Chen}},
  \bibinfo{author}{\bibfnamefont{E.}~\bibnamefont{Diez}},
  \bibinfo{author}{\bibfnamefont{D.~L.} \bibnamefont{Sivco}}, \bibnamefont{and}
  \bibinfo{author}{\bibfnamefont{A.~Y.} \bibnamefont{Cho}},
  \bibinfo{journal}{Physical Review Letters} \textbf{\bibinfo{volume}{91}},
  \bibinfo{pages}{236802} (\bibinfo{year}{2003}),
  \urlprefix\url{http://link.aps.org/doi/10.1103/PhysRevLett.91.236802}.

\bibitem[{\citenamefont{Staring et~al.}(1992)\citenamefont{Staring, van Houten,
  Beenakker, and Foxon}}]{staring_coulomb-blockade_1992}
\bibinfo{author}{\bibfnamefont{A.~A.~M.} \bibnamefont{Staring}},
  \bibinfo{author}{\bibfnamefont{H.}~\bibnamefont{van Houten}},
  \bibinfo{author}{\bibfnamefont{C.~W.~J.} \bibnamefont{Beenakker}},
  \bibnamefont{and} \bibinfo{author}{\bibfnamefont{C.~T.} \bibnamefont{Foxon}},
  \bibinfo{journal}{Physical Review B} \textbf{\bibinfo{volume}{45}},
  \bibinfo{pages}{9222} (\bibinfo{year}{1992}),
  \urlprefix\url{http://link.aps.org/doi/10.1103/PhysRevB.45.9222}.

\bibitem[{\citenamefont{Main et~al.}(1994)\citenamefont{Main, Geim, Carmona,
  Brown, Foster, Taboryski, and Lindelof}}]{main_resistance_1994}
\bibinfo{author}{\bibfnamefont{P.~C.} \bibnamefont{Main}},
  \bibinfo{author}{\bibfnamefont{A.~K.} \bibnamefont{Geim}},
  \bibinfo{author}{\bibfnamefont{H.~A.} \bibnamefont{Carmona}},
  \bibinfo{author}{\bibfnamefont{C.~V.} \bibnamefont{Brown}},
  \bibinfo{author}{\bibfnamefont{T.~J.} \bibnamefont{Foster}},
  \bibinfo{author}{\bibfnamefont{R.}~\bibnamefont{Taboryski}},
  \bibnamefont{and} \bibinfo{author}{\bibfnamefont{P.~E.}
  \bibnamefont{Lindelof}}, \bibinfo{journal}{Physical Review B}
  \textbf{\bibinfo{volume}{50}}, \bibinfo{pages}{4450} (\bibinfo{year}{1994}),
  \urlprefix\url{http://link.aps.org/doi/10.1103/PhysRevB.50.4450}.

\bibitem[{\citenamefont{Timp et~al.}(1987)\citenamefont{Timp, Chang,
  Mankiewich, Behringer, Cunningham, Chang, and Howard}}]{timp_quantum_1987}
\bibinfo{author}{\bibfnamefont{G.}~\bibnamefont{Timp}},
  \bibinfo{author}{\bibfnamefont{A.~M.} \bibnamefont{Chang}},
  \bibinfo{author}{\bibfnamefont{P.}~\bibnamefont{Mankiewich}},
  \bibinfo{author}{\bibfnamefont{R.}~\bibnamefont{Behringer}},
  \bibinfo{author}{\bibfnamefont{J.~E.} \bibnamefont{Cunningham}},
  \bibinfo{author}{\bibfnamefont{T.~Y.} \bibnamefont{Chang}}, \bibnamefont{and}
  \bibinfo{author}{\bibfnamefont{R.~E.} \bibnamefont{Howard}},
  \bibinfo{journal}{Physical Review Letters} \textbf{\bibinfo{volume}{59}},
  \bibinfo{pages}{732} (\bibinfo{year}{1987}),
  \urlprefix\url{http://link.aps.org/doi/10.1103/PhysRevLett.59.732}.

\bibitem[{\citenamefont{Velasco et~al.}(2010)\citenamefont{Velasco, Liu, Jing,
  Kratz, Zhang, Bao, Bockrath, and Lau}}]{velasco_probing_2010}
\bibinfo{author}{\bibfnamefont{J.}~\bibnamefont{Velasco}},
  \bibinfo{author}{\bibfnamefont{G.}~\bibnamefont{Liu}},
  \bibinfo{author}{\bibfnamefont{L.}~\bibnamefont{Jing}},
  \bibinfo{author}{\bibfnamefont{P.}~\bibnamefont{Kratz}},
  \bibinfo{author}{\bibfnamefont{H.}~\bibnamefont{Zhang}},
  \bibinfo{author}{\bibfnamefont{W.}~\bibnamefont{Bao}},
  \bibinfo{author}{\bibfnamefont{M.}~\bibnamefont{Bockrath}}, \bibnamefont{and}
  \bibinfo{author}{\bibfnamefont{C.~N.} \bibnamefont{Lau}},
  \bibinfo{journal}{Physical Review B} \textbf{\bibinfo{volume}{81}},
  \bibinfo{pages}{121407} (\bibinfo{year}{2010}),
  \urlprefix\url{http://link.aps.org/doi/10.1103/PhysRevB.81.121407}.

\bibitem[{\citenamefont{Martin et~al.}(2009)\citenamefont{Martin, Akerman,
  Ulbricht, Lohmann, von Klitzing, Smet, and Yacoby}}]{martin_nature_2009}
\bibinfo{author}{\bibfnamefont{J.}~\bibnamefont{Martin}},
  \bibinfo{author}{\bibfnamefont{N.}~\bibnamefont{Akerman}},
  \bibinfo{author}{\bibfnamefont{G.}~\bibnamefont{Ulbricht}},
  \bibinfo{author}{\bibfnamefont{T.}~\bibnamefont{Lohmann}},
  \bibinfo{author}{\bibfnamefont{K.}~\bibnamefont{von Klitzing}},
  \bibinfo{author}{\bibfnamefont{J.~H.} \bibnamefont{Smet}}, \bibnamefont{and}
  \bibinfo{author}{\bibfnamefont{A.}~\bibnamefont{Yacoby}},
  \bibinfo{journal}{Nature Physics} \textbf{\bibinfo{volume}{5}},
  \bibinfo{pages}{669} (\bibinfo{year}{2009}), ISSN \bibinfo{issn}{1745-2473},
  \urlprefix\url{http://www.nature.com/nphys/journal/v5/n9/full/nphys1344.html}.

\bibitem[{\citenamefont{Branchaud et~al.}(2010)\citenamefont{Branchaud, Kam,
  Zawadzki, Peeters, and Sachrajda}}]{branchaud_transport_2010}
\bibinfo{author}{\bibfnamefont{S.}~\bibnamefont{Branchaud}},
  \bibinfo{author}{\bibfnamefont{A.}~\bibnamefont{Kam}},
  \bibinfo{author}{\bibfnamefont{P.}~\bibnamefont{Zawadzki}},
  \bibinfo{author}{\bibfnamefont{F.~M.} \bibnamefont{Peeters}},
  \bibnamefont{and} \bibinfo{author}{\bibfnamefont{A.~S.}
  \bibnamefont{Sachrajda}}, \bibinfo{journal}{Physical Review B}
  \textbf{\bibinfo{volume}{81}}, \bibinfo{pages}{121406}
  (\bibinfo{year}{2010}),
  \urlprefix\url{http://link.aps.org/doi/10.1103/PhysRevB.81.121406}.

\bibitem[{\citenamefont{Granger et~al.}(2011)\citenamefont{Granger, Studenikin,
  Kam, Sachrajda, and Poole}}]{granger_few-electron_2011}
\bibinfo{author}{\bibfnamefont{G.}~\bibnamefont{Granger}},
  \bibinfo{author}{\bibfnamefont{S.~A.} \bibnamefont{Studenikin}},
  \bibinfo{author}{\bibfnamefont{A.}~\bibnamefont{Kam}},
  \bibinfo{author}{\bibfnamefont{A.~S.} \bibnamefont{Sachrajda}},
  \bibnamefont{and} \bibinfo{author}{\bibfnamefont{P.~J.} \bibnamefont{Poole}},
  \bibinfo{journal}{Applied Physics Letters} \textbf{\bibinfo{volume}{98}},
  \bibinfo{pages}{132107} (\bibinfo{year}{2011}), ISSN
  \bibinfo{issn}{0003-6951, 1077-3118},
  \urlprefix\url{http://scitation.aip.org/content/aip/journal/apl/98/13/10.1063/1.3574540}.

\bibitem[{\citenamefont{Lee}(1990)}]{lee_comment_1990}
\bibinfo{author}{\bibfnamefont{P.~A.} \bibnamefont{Lee}},
  \bibinfo{journal}{Physical Review Letters} \textbf{\bibinfo{volume}{65}},
  \bibinfo{pages}{2206} (\bibinfo{year}{1990}),
  \urlprefix\url{http://link.aps.org/doi/10.1103/PhysRevLett.65.2206}.

\bibitem[{\citenamefont{Rosenow and Halperin}(2007)}]{rosenow_influence_2007}
\bibinfo{author}{\bibfnamefont{B.}~\bibnamefont{Rosenow}} \bibnamefont{and}
  \bibinfo{author}{\bibfnamefont{B.~I.} \bibnamefont{Halperin}},
  \bibinfo{journal}{Physical Review Letters} \textbf{\bibinfo{volume}{98}},
  \bibinfo{pages}{106801} (\bibinfo{year}{2007}),
  \urlprefix\url{http://link.aps.org/doi/10.1103/PhysRevLett.98.106801}.

\bibitem[{\citenamefont{Halperin et~al.}(2011)\citenamefont{Halperin, Stern,
  Neder, and Rosenow}}]{halperin_theory_2011}
\bibinfo{author}{\bibfnamefont{B.~I.} \bibnamefont{Halperin}},
  \bibinfo{author}{\bibfnamefont{A.}~\bibnamefont{Stern}},
  \bibinfo{author}{\bibfnamefont{I.}~\bibnamefont{Neder}}, \bibnamefont{and}
  \bibinfo{author}{\bibfnamefont{B.}~\bibnamefont{Rosenow}},
  \bibinfo{journal}{Physical Review B} \textbf{\bibinfo{volume}{83}},
  \bibinfo{pages}{155440} (\bibinfo{year}{2011}),
  \urlprefix\url{http://link.aps.org/doi/10.1103/PhysRevB.83.155440}.

\bibitem[{\citenamefont{Hackens et~al.}(2010)\citenamefont{Hackens, Martins,
  Faniel, Dutu, Sellier, Huant, Pala, Desplanque, Wallart, and
  Bayot}}]{hackens_imaging_2010}
\bibinfo{author}{\bibfnamefont{B.}~\bibnamefont{Hackens}},
  \bibinfo{author}{\bibfnamefont{F.}~\bibnamefont{Martins}},
  \bibinfo{author}{\bibfnamefont{S.}~\bibnamefont{Faniel}},
  \bibinfo{author}{\bibfnamefont{C.~A.} \bibnamefont{Dutu}},
  \bibinfo{author}{\bibfnamefont{H.}~\bibnamefont{Sellier}},
  \bibinfo{author}{\bibfnamefont{S.}~\bibnamefont{Huant}},
  \bibinfo{author}{\bibfnamefont{M.}~\bibnamefont{Pala}},
  \bibinfo{author}{\bibfnamefont{L.}~\bibnamefont{Desplanque}},
  \bibinfo{author}{\bibfnamefont{X.}~\bibnamefont{Wallart}}, \bibnamefont{and}
  \bibinfo{author}{\bibfnamefont{V.}~\bibnamefont{Bayot}},
  \bibinfo{journal}{Nature Communications} \textbf{\bibinfo{volume}{1}},
  \bibinfo{pages}{39} (\bibinfo{year}{2010}),
  \urlprefix\url{http://www.nature.com/ncomms/journal/v1/n4/abs/ncomms1038.html}.

\bibitem[{\citenamefont{Martins
  et~al.}(2013{\natexlab{a}})\citenamefont{Martins, Faniel, Rosenow, Pala,
  Sellier, Huant, Desplanque, Wallart, Bayot, and
  Hackens}}]{martins_scanning_2013}
\bibinfo{author}{\bibfnamefont{F.}~\bibnamefont{Martins}},
  \bibinfo{author}{\bibfnamefont{S.}~\bibnamefont{Faniel}},
  \bibinfo{author}{\bibfnamefont{B.}~\bibnamefont{Rosenow}},
  \bibinfo{author}{\bibfnamefont{M.~G.} \bibnamefont{Pala}},
  \bibinfo{author}{\bibfnamefont{H.}~\bibnamefont{Sellier}},
  \bibinfo{author}{\bibfnamefont{S.}~\bibnamefont{Huant}},
  \bibinfo{author}{\bibfnamefont{L.}~\bibnamefont{Desplanque}},
  \bibinfo{author}{\bibfnamefont{X.}~\bibnamefont{Wallart}},
  \bibinfo{author}{\bibfnamefont{V.}~\bibnamefont{Bayot}}, \bibnamefont{and}
  \bibinfo{author}{\bibfnamefont{B.}~\bibnamefont{Hackens}},
  \bibinfo{journal}{New Journal of Physics} \textbf{\bibinfo{volume}{15}},
  \bibinfo{pages}{013049} (\bibinfo{year}{2013}{\natexlab{a}}), ISSN
  \bibinfo{issn}{1367-2630},
  \urlprefix\url{http://iopscience.iop.org/1367-2630/15/1/013049}.

\bibitem[{\citenamefont{Martins
  et~al.}(2013{\natexlab{b}})\citenamefont{Martins, Faniel, Rosenow, Sellier,
  Huant, Pala, Desplanque, Wallart, Bayot, and
  Hackens}}]{martins_coherent_2013}
\bibinfo{author}{\bibfnamefont{F.}~\bibnamefont{Martins}},
  \bibinfo{author}{\bibfnamefont{S.}~\bibnamefont{Faniel}},
  \bibinfo{author}{\bibfnamefont{B.}~\bibnamefont{Rosenow}},
  \bibinfo{author}{\bibfnamefont{H.}~\bibnamefont{Sellier}},
  \bibinfo{author}{\bibfnamefont{S.}~\bibnamefont{Huant}},
  \bibinfo{author}{\bibfnamefont{M.~G.} \bibnamefont{Pala}},
  \bibinfo{author}{\bibfnamefont{L.}~\bibnamefont{Desplanque}},
  \bibinfo{author}{\bibfnamefont{X.}~\bibnamefont{Wallart}},
  \bibinfo{author}{\bibfnamefont{V.}~\bibnamefont{Bayot}}, \bibnamefont{and}
  \bibinfo{author}{\bibfnamefont{B.}~\bibnamefont{Hackens}},
  \bibinfo{journal}{Scientific Reports} \textbf{\bibinfo{volume}{3}}
  (\bibinfo{year}{2013}{\natexlab{b}}),
  \urlprefix\url{http://www.nature.com/srep/2013/130311/srep01416/full/srep01416.html}.

\bibitem[{\citenamefont{Martin et~al.}(2008)\citenamefont{Martin, Szorkovszky,
  Micolich, Hamilton, Marlow, Linke, Taylor, and
  Samuelson}}]{martin_enhanced_2008}
\bibinfo{author}{\bibfnamefont{T.~P.} \bibnamefont{Martin}},
  \bibinfo{author}{\bibfnamefont{A.}~\bibnamefont{Szorkovszky}},
  \bibinfo{author}{\bibfnamefont{A.~P.} \bibnamefont{Micolich}},
  \bibinfo{author}{\bibfnamefont{A.~R.} \bibnamefont{Hamilton}},
  \bibinfo{author}{\bibfnamefont{C.~A.} \bibnamefont{Marlow}},
  \bibinfo{author}{\bibfnamefont{H.}~\bibnamefont{Linke}},
  \bibinfo{author}{\bibfnamefont{R.~P.} \bibnamefont{Taylor}},
  \bibnamefont{and}
  \bibinfo{author}{\bibfnamefont{L.}~\bibnamefont{Samuelson}},
  \bibinfo{journal}{Applied Physics Letters} \textbf{\bibinfo{volume}{93}},
  \bibinfo{pages}{012105} (\bibinfo{year}{2008}), ISSN
  \bibinfo{issn}{00036951},
  \urlprefix\url{http://apl.aip.org/resource/1/applab/v93/i1/p012105_s1}.

\bibitem[{\citenamefont{Yoon et~al.}(2009)\citenamefont{Yoon, Kang, Ivanushkin,
  Mourokh, Morimoto, Aoki, Reno, Ochiai, and Bird}}]{yoon_nonlocal_2009}
\bibinfo{author}{\bibfnamefont{Y.}~\bibnamefont{Yoon}},
  \bibinfo{author}{\bibfnamefont{M.-G.} \bibnamefont{Kang}},
  \bibinfo{author}{\bibfnamefont{P.}~\bibnamefont{Ivanushkin}},
  \bibinfo{author}{\bibfnamefont{L.}~\bibnamefont{Mourokh}},
  \bibinfo{author}{\bibfnamefont{T.}~\bibnamefont{Morimoto}},
  \bibinfo{author}{\bibfnamefont{N.}~\bibnamefont{Aoki}},
  \bibinfo{author}{\bibfnamefont{J.~L.} \bibnamefont{Reno}},
  \bibinfo{author}{\bibfnamefont{Y.}~\bibnamefont{Ochiai}}, \bibnamefont{and}
  \bibinfo{author}{\bibfnamefont{J.~P.} \bibnamefont{Bird}},
  \bibinfo{journal}{Applied Physics Letters} \textbf{\bibinfo{volume}{94}},
  \bibinfo{pages}{213103} (\bibinfo{year}{2009}), ISSN
  \bibinfo{issn}{00036951},
  \urlprefix\url{http://apl.aip.org/resource/1/applab/v94/i21/p213103_s1}.

\bibitem[{\citenamefont{R{\"o}ssler et~al.}(2011)\citenamefont{R{\"o}ssler,
  Baer, de~Wiljes, Ardelt, Ihn, Ensslin, Reichl, and
  Wegscheider}}]{rossler_transport_2011}
\bibinfo{author}{\bibfnamefont{C.}~\bibnamefont{R{\"o}ssler}},
  \bibinfo{author}{\bibfnamefont{S.}~\bibnamefont{Baer}},
  \bibinfo{author}{\bibfnamefont{E.}~\bibnamefont{de~Wiljes}},
  \bibinfo{author}{\bibfnamefont{P.-L.} \bibnamefont{Ardelt}},
  \bibinfo{author}{\bibfnamefont{T.}~\bibnamefont{Ihn}},
  \bibinfo{author}{\bibfnamefont{K.}~\bibnamefont{Ensslin}},
  \bibinfo{author}{\bibfnamefont{C.}~\bibnamefont{Reichl}}, \bibnamefont{and}
  \bibinfo{author}{\bibfnamefont{W.}~\bibnamefont{Wegscheider}},
  \bibinfo{journal}{New Journal of Physics} \textbf{\bibinfo{volume}{13}},
  \bibinfo{pages}{113006} (\bibinfo{year}{2011}), ISSN
  \bibinfo{issn}{1367-2630},
  \urlprefix\url{http://iopscience.iop.org/1367-2630/13/11/113006}.

\bibitem[{\citenamefont{Thomas et~al.}(1996)\citenamefont{Thomas, Nicholls,
  Simmons, Pepper, Mace, and Ritchie}}]{thomas_possible_1996}
\bibinfo{author}{\bibfnamefont{K.~J.} \bibnamefont{Thomas}},
  \bibinfo{author}{\bibfnamefont{J.~T.} \bibnamefont{Nicholls}},
  \bibinfo{author}{\bibfnamefont{M.~Y.} \bibnamefont{Simmons}},
  \bibinfo{author}{\bibfnamefont{M.}~\bibnamefont{Pepper}},
  \bibinfo{author}{\bibfnamefont{D.~R.} \bibnamefont{Mace}}, \bibnamefont{and}
  \bibinfo{author}{\bibfnamefont{D.~A.} \bibnamefont{Ritchie}},
  \bibinfo{journal}{Physical Review Letters} \textbf{\bibinfo{volume}{77}},
  \bibinfo{pages}{135} (\bibinfo{year}{1996}),
  \urlprefix\url{http://link.aps.org/doi/10.1103/PhysRevLett.77.135}.

\bibitem[{\citenamefont{Kristensen et~al.}(2000)\citenamefont{Kristensen,
  Bruus, Hansen, Jensen, Lindelof, Marckmann, Nyg{\r a}rd, S{\o}rensen,
  Beuscher, Forchel et~al.}}]{kristensen_bias_2000}
\bibinfo{author}{\bibfnamefont{A.}~\bibnamefont{Kristensen}},
  \bibinfo{author}{\bibfnamefont{H.}~\bibnamefont{Bruus}},
  \bibinfo{author}{\bibfnamefont{A.~E.} \bibnamefont{Hansen}},
  \bibinfo{author}{\bibfnamefont{J.~B.} \bibnamefont{Jensen}},
  \bibinfo{author}{\bibfnamefont{P.~E.} \bibnamefont{Lindelof}},
  \bibinfo{author}{\bibfnamefont{C.~J.} \bibnamefont{Marckmann}},
  \bibinfo{author}{\bibfnamefont{J.}~\bibnamefont{Nyg{\r a}rd}},
  \bibinfo{author}{\bibfnamefont{C.~B.} \bibnamefont{S{\o}rensen}},
  \bibinfo{author}{\bibfnamefont{F.}~\bibnamefont{Beuscher}},
  \bibinfo{author}{\bibfnamefont{A.}~\bibnamefont{Forchel}},
  \bibnamefont{et~al.}, \bibinfo{journal}{Physical Review B}
  \textbf{\bibinfo{volume}{62}}, \bibinfo{pages}{10950} (\bibinfo{year}{2000}),
  \urlprefix\url{http://link.aps.org/doi/10.1103/PhysRevB.62.10950}.

\bibitem[{\citenamefont{Cronenwett et~al.}(2002)\citenamefont{Cronenwett,
  Lynch, Goldhaber-Gordon, Kouwenhoven, Marcus, Hirose, Wingreen, and
  Umansky}}]{cronenwett_low-temperature_2002}
\bibinfo{author}{\bibfnamefont{S.~M.} \bibnamefont{Cronenwett}},
  \bibinfo{author}{\bibfnamefont{H.~J.} \bibnamefont{Lynch}},
  \bibinfo{author}{\bibfnamefont{D.}~\bibnamefont{Goldhaber-Gordon}},
  \bibinfo{author}{\bibfnamefont{L.~P.} \bibnamefont{Kouwenhoven}},
  \bibinfo{author}{\bibfnamefont{C.~M.} \bibnamefont{Marcus}},
  \bibinfo{author}{\bibfnamefont{K.}~\bibnamefont{Hirose}},
  \bibinfo{author}{\bibfnamefont{N.~S.} \bibnamefont{Wingreen}},
  \bibnamefont{and} \bibinfo{author}{\bibfnamefont{V.}~\bibnamefont{Umansky}},
  \bibinfo{journal}{Physical Review Letters} \textbf{\bibinfo{volume}{88}},
  \bibinfo{pages}{226805} (\bibinfo{year}{2002}),
  \urlprefix\url{http://link.aps.org/doi/10.1103/PhysRevLett.88.226805}.

\bibitem[{\citenamefont{Komijani et~al.}(2010)\citenamefont{Komijani, Csontos,
  Shorubalko, Ihn, Ensslin, Meir, Reuter, and Wieck}}]{komijani_evidence_2010}
\bibinfo{author}{\bibfnamefont{Y.}~\bibnamefont{Komijani}},
  \bibinfo{author}{\bibfnamefont{M.}~\bibnamefont{Csontos}},
  \bibinfo{author}{\bibfnamefont{I.}~\bibnamefont{Shorubalko}},
  \bibinfo{author}{\bibfnamefont{T.}~\bibnamefont{Ihn}},
  \bibinfo{author}{\bibfnamefont{K.}~\bibnamefont{Ensslin}},
  \bibinfo{author}{\bibfnamefont{Y.}~\bibnamefont{Meir}},
  \bibinfo{author}{\bibfnamefont{D.}~\bibnamefont{Reuter}}, \bibnamefont{and}
  \bibinfo{author}{\bibfnamefont{A.~D.} \bibnamefont{Wieck}},
  \bibinfo{journal}{{EPL} (Europhysics Letters)} \textbf{\bibinfo{volume}{91}},
  \bibinfo{pages}{67010} (\bibinfo{year}{2010}), ISSN
  \bibinfo{issn}{0295-5075},
  \urlprefix\url{http://iopscience.iop.org/0295-5075/91/6/67010}.

\bibitem[{\citenamefont{Micolich}(2011)}]{micolich_what_2011}
\bibinfo{author}{\bibfnamefont{A.~P.} \bibnamefont{Micolich}},
  \bibinfo{journal}{Journal of Physics: Condensed Matter}
  \textbf{\bibinfo{volume}{23}}, \bibinfo{pages}{443201}
  (\bibinfo{year}{2011}), ISSN \bibinfo{issn}{0953-8984},
  \urlprefix\url{http://iopscience.iop.org/0953-8984/23/44/443201}.

\bibitem[{\citenamefont{Komijani et~al.}(2013)\citenamefont{Komijani, Csontos,
  Ihn, Ensslin, Meir, Reuter, and Wieck}}]{komijani_origins_2013}
\bibinfo{author}{\bibfnamefont{Y.}~\bibnamefont{Komijani}},
  \bibinfo{author}{\bibfnamefont{M.}~\bibnamefont{Csontos}},
  \bibinfo{author}{\bibfnamefont{T.}~\bibnamefont{Ihn}},
  \bibinfo{author}{\bibfnamefont{K.}~\bibnamefont{Ensslin}},
  \bibinfo{author}{\bibfnamefont{Y.}~\bibnamefont{Meir}},
  \bibinfo{author}{\bibfnamefont{D.}~\bibnamefont{Reuter}}, \bibnamefont{and}
  \bibinfo{author}{\bibfnamefont{A.~D.} \bibnamefont{Wieck}},
  \bibinfo{journal}{Physical Review B} \textbf{\bibinfo{volume}{87}},
  \bibinfo{pages}{245406} (\bibinfo{year}{2013}),
  \urlprefix\url{http://link.aps.org/doi/10.1103/PhysRevB.87.245406}.

\bibitem[{\citenamefont{Gustavsson et~al.}(2006)\citenamefont{Gustavsson,
  Leturcq, Simovi{\v c}, Schleser, Ihn, Studerus, Ensslin, Driscoll, and
  Gossard}}]{gustavsson_counting_2006-1}
\bibinfo{author}{\bibfnamefont{S.}~\bibnamefont{Gustavsson}},
  \bibinfo{author}{\bibfnamefont{R.}~\bibnamefont{Leturcq}},
  \bibinfo{author}{\bibfnamefont{B.}~\bibnamefont{Simovi{\v c}}},
  \bibinfo{author}{\bibfnamefont{R.}~\bibnamefont{Schleser}},
  \bibinfo{author}{\bibfnamefont{T.}~\bibnamefont{Ihn}},
  \bibinfo{author}{\bibfnamefont{P.}~\bibnamefont{Studerus}},
  \bibinfo{author}{\bibfnamefont{K.}~\bibnamefont{Ensslin}},
  \bibinfo{author}{\bibfnamefont{D.~C.} \bibnamefont{Driscoll}},
  \bibnamefont{and} \bibinfo{author}{\bibfnamefont{A.~C.}
  \bibnamefont{Gossard}}, \bibinfo{journal}{Physical Review Letters}
  \textbf{\bibinfo{volume}{96}}, \bibinfo{pages}{076605}
  (\bibinfo{year}{2006}),
  \urlprefix\url{http://link.aps.org/doi/10.1103/PhysRevLett.96.076605}.

\bibitem[{\citenamefont{Gustavsson et~al.}(2007)\citenamefont{Gustavsson,
  Studer, Leturcq, Ihn, Ensslin, Driscoll, and
  Gossard}}]{gustavsson_frequency-selective_2007}
\bibinfo{author}{\bibfnamefont{S.}~\bibnamefont{Gustavsson}},
  \bibinfo{author}{\bibfnamefont{M.}~\bibnamefont{Studer}},
  \bibinfo{author}{\bibfnamefont{R.}~\bibnamefont{Leturcq}},
  \bibinfo{author}{\bibfnamefont{T.}~\bibnamefont{Ihn}},
  \bibinfo{author}{\bibfnamefont{K.}~\bibnamefont{Ensslin}},
  \bibinfo{author}{\bibfnamefont{D.~C.} \bibnamefont{Driscoll}},
  \bibnamefont{and} \bibinfo{author}{\bibfnamefont{A.~C.}
  \bibnamefont{Gossard}}, \bibinfo{journal}{Physical Review Letters}
  \textbf{\bibinfo{volume}{99}}, \bibinfo{pages}{206804}
  (\bibinfo{year}{2007}),
  \urlprefix\url{http://link.aps.org/doi/10.1103/PhysRevLett.99.206804}.

\bibitem[{\citenamefont{R{\"o}ssler et~al.}(2013)\citenamefont{R{\"o}ssler,
  Kr{\"a}henmann, Baer, Ihn, Ensslin, Reichl, and
  Wegscheider}}]{rossler_tunable_2013}
\bibinfo{author}{\bibfnamefont{C.}~\bibnamefont{R{\"o}ssler}},
  \bibinfo{author}{\bibfnamefont{T.}~\bibnamefont{Kr{\"a}henmann}},
  \bibinfo{author}{\bibfnamefont{S.}~\bibnamefont{Baer}},
  \bibinfo{author}{\bibfnamefont{T.}~\bibnamefont{Ihn}},
  \bibinfo{author}{\bibfnamefont{K.}~\bibnamefont{Ensslin}},
  \bibinfo{author}{\bibfnamefont{C.}~\bibnamefont{Reichl}}, \bibnamefont{and}
  \bibinfo{author}{\bibfnamefont{W.}~\bibnamefont{Wegscheider}},
  \bibinfo{journal}{New Journal of Physics} \textbf{\bibinfo{volume}{15}},
  \bibinfo{pages}{033011} (\bibinfo{year}{2013}), ISSN
  \bibinfo{issn}{1367-2630},
  \urlprefix\url{http://iopscience.iop.org/1367-2630/15/3/033011}.

\bibitem[{\citenamefont{Baer et~al.}(2013)\citenamefont{Baer, R{\"o}ssler, Ihn,
  Ensslin, Reichl, and Wegscheider}}]{baer_cyclic_2013}
\bibinfo{author}{\bibfnamefont{S.}~\bibnamefont{Baer}},
  \bibinfo{author}{\bibfnamefont{C.}~\bibnamefont{R{\"o}ssler}},
  \bibinfo{author}{\bibfnamefont{T.}~\bibnamefont{Ihn}},
  \bibinfo{author}{\bibfnamefont{K.}~\bibnamefont{Ensslin}},
  \bibinfo{author}{\bibfnamefont{C.}~\bibnamefont{Reichl}}, \bibnamefont{and}
  \bibinfo{author}{\bibfnamefont{W.}~\bibnamefont{Wegscheider}},
  \bibinfo{journal}{New Journal of Physics} \textbf{\bibinfo{volume}{15}},
  \bibinfo{pages}{023035} (\bibinfo{year}{2013}), ISSN
  \bibinfo{issn}{1367-2630},
  \urlprefix\url{http://iopscience.iop.org/1367-2630/15/2/023035}.

\bibitem[{\citenamefont{Milliken et~al.}(1996)\citenamefont{Milliken, Umbach,
  and Webb}}]{milliken_indications_1996}
\bibinfo{author}{\bibfnamefont{F.}~\bibnamefont{Milliken}},
  \bibinfo{author}{\bibfnamefont{C.}~\bibnamefont{Umbach}}, \bibnamefont{and}
  \bibinfo{author}{\bibfnamefont{R.}~\bibnamefont{Webb}},
  \bibinfo{journal}{Solid State Communications} \textbf{\bibinfo{volume}{97}},
  \bibinfo{pages}{309} (\bibinfo{year}{1996}), ISSN \bibinfo{issn}{0038-1098},
  \urlprefix\url{http://www.sciencedirect.com/science/article/pii/0038109895001816}.

\bibitem[{\citenamefont{Chang et~al.}(1996)\citenamefont{Chang, Pfeiffer, and
  West}}]{chang_observation_1996}
\bibinfo{author}{\bibfnamefont{A.~M.} \bibnamefont{Chang}},
  \bibinfo{author}{\bibfnamefont{L.~N.} \bibnamefont{Pfeiffer}},
  \bibnamefont{and} \bibinfo{author}{\bibfnamefont{K.~W.} \bibnamefont{West}},
  \bibinfo{journal}{Physical Review Letters} \textbf{\bibinfo{volume}{77}},
  \bibinfo{pages}{2538} (\bibinfo{year}{1996}),
  \urlprefix\url{http://link.aps.org/doi/10.1103/PhysRevLett.77.2538}.

\bibitem[{\citenamefont{Roddaro et~al.}(2004)\citenamefont{Roddaro, Pellegrini,
  Beltram, Biasiol, and Sorba}}]{roddaro_interedge_2004}
\bibinfo{author}{\bibfnamefont{S.}~\bibnamefont{Roddaro}},
  \bibinfo{author}{\bibfnamefont{V.}~\bibnamefont{Pellegrini}},
  \bibinfo{author}{\bibfnamefont{F.}~\bibnamefont{Beltram}},
  \bibinfo{author}{\bibfnamefont{G.}~\bibnamefont{Biasiol}}, \bibnamefont{and}
  \bibinfo{author}{\bibfnamefont{L.}~\bibnamefont{Sorba}},
  \bibinfo{journal}{Physical Review Letters} \textbf{\bibinfo{volume}{93}},
  \bibinfo{pages}{046801} (\bibinfo{year}{2004}),
  \urlprefix\url{http://link.aps.org/doi/10.1103/PhysRevLett.93.046801}.

\bibitem[{\citenamefont{Zhang et~al.}(2009)\citenamefont{Zhang, {McClure},
  Levenson-Falk, Marcus, Pfeiffer, and West}}]{zhang_distinct_2009}
\bibinfo{author}{\bibfnamefont{Y.}~\bibnamefont{Zhang}},
  \bibinfo{author}{\bibfnamefont{D.~T.} \bibnamefont{{McClure}}},
  \bibinfo{author}{\bibfnamefont{E.~M.} \bibnamefont{Levenson-Falk}},
  \bibinfo{author}{\bibfnamefont{C.~M.} \bibnamefont{Marcus}},
  \bibinfo{author}{\bibfnamefont{L.~N.} \bibnamefont{Pfeiffer}},
  \bibnamefont{and} \bibinfo{author}{\bibfnamefont{K.~W.} \bibnamefont{West}},
  \bibinfo{journal}{Physical Review B} \textbf{\bibinfo{volume}{79}},
  \bibinfo{pages}{241304} (\bibinfo{year}{2009}),
  \urlprefix\url{http://link.aps.org/doi/10.1103/PhysRevB.79.241304}.

\bibitem[{\citenamefont{Kou et~al.}(2012)\citenamefont{Kou, Marcus, Pfeiffer,
  and West}}]{kou_coulomb_2012}
\bibinfo{author}{\bibfnamefont{A.}~\bibnamefont{Kou}},
  \bibinfo{author}{\bibfnamefont{C.~M.} \bibnamefont{Marcus}},
  \bibinfo{author}{\bibfnamefont{L.~N.} \bibnamefont{Pfeiffer}},
  \bibnamefont{and} \bibinfo{author}{\bibfnamefont{K.~W.} \bibnamefont{West}},
  \bibinfo{journal}{Physical Review Letters} \textbf{\bibinfo{volume}{108}},
  \bibinfo{pages}{256803} (\bibinfo{year}{2012}),
  \urlprefix\url{http://link.aps.org/doi/10.1103/PhysRevLett.108.256803}.

\bibitem[{\citenamefont{Camino et~al.}(2007)\citenamefont{Camino, Zhou, and
  Goldman}}]{camino_quantum_2007}
\bibinfo{author}{\bibfnamefont{F.~E.} \bibnamefont{Camino}},
  \bibinfo{author}{\bibfnamefont{W.}~\bibnamefont{Zhou}}, \bibnamefont{and}
  \bibinfo{author}{\bibfnamefont{V.~J.} \bibnamefont{Goldman}},
  \bibinfo{journal}{Physical Review B} \textbf{\bibinfo{volume}{76}},
  \bibinfo{pages}{155305} (\bibinfo{year}{2007}),
  \urlprefix\url{http://link.aps.org/doi/10.1103/PhysRevB.76.155305}.

\bibitem[{\citenamefont{{McClure} et~al.}(2012)\citenamefont{{McClure}, Chang,
  Marcus, Pfeiffer, and West}}]{mcclure_fabry-perot_2012}
\bibinfo{author}{\bibfnamefont{D.~T.} \bibnamefont{{McClure}}},
  \bibinfo{author}{\bibfnamefont{W.}~\bibnamefont{Chang}},
  \bibinfo{author}{\bibfnamefont{C.~M.} \bibnamefont{Marcus}},
  \bibinfo{author}{\bibfnamefont{L.~N.} \bibnamefont{Pfeiffer}},
  \bibnamefont{and} \bibinfo{author}{\bibfnamefont{K.~W.} \bibnamefont{West}},
  \bibinfo{journal}{Physical Review Letters} \textbf{\bibinfo{volume}{108}},
  \bibinfo{pages}{256804} (\bibinfo{year}{2012}),
  \urlprefix\url{http://link.aps.org/doi/10.1103/PhysRevLett.108.256804}.

\bibitem[{\citenamefont{Willett et~al.}(2009)\citenamefont{Willett, Pfeiffer,
  and West}}]{willett_measurement_2009}
\bibinfo{author}{\bibfnamefont{R.~L.} \bibnamefont{Willett}},
  \bibinfo{author}{\bibfnamefont{L.~N.} \bibnamefont{Pfeiffer}},
  \bibnamefont{and} \bibinfo{author}{\bibfnamefont{K.~W.} \bibnamefont{West}},
  \bibinfo{journal}{Proceedings of the National Academy of Sciences}
  \textbf{\bibinfo{volume}{106}}, \bibinfo{pages}{8853} (\bibinfo{year}{2009}),
  ISSN \bibinfo{issn}{0027-8424, 1091-6490},
  \urlprefix\url{http://www.pnas.org/content/106/22/8853}.

\bibitem[{\citenamefont{R{\"o}ssler et~al.}(2010)\citenamefont{R{\"o}ssler,
  Feil, Mensch, Ihn, Ensslin, Schuh, and Wegscheider}}]{rossler_gating_2010}
\bibinfo{author}{\bibfnamefont{C.}~\bibnamefont{R{\"o}ssler}},
  \bibinfo{author}{\bibfnamefont{T.}~\bibnamefont{Feil}},
  \bibinfo{author}{\bibfnamefont{P.}~\bibnamefont{Mensch}},
  \bibinfo{author}{\bibfnamefont{T.}~\bibnamefont{Ihn}},
  \bibinfo{author}{\bibfnamefont{K.}~\bibnamefont{Ensslin}},
  \bibinfo{author}{\bibfnamefont{D.}~\bibnamefont{Schuh}}, \bibnamefont{and}
  \bibinfo{author}{\bibfnamefont{W.}~\bibnamefont{Wegscheider}},
  \bibinfo{journal}{New Journal of Physics} \textbf{\bibinfo{volume}{12}},
  \bibinfo{pages}{043007} (\bibinfo{year}{2010}), ISSN
  \bibinfo{issn}{1367-2630},
  \urlprefix\url{http://iopscience.iop.org/1367-2630/12/4/043007}.

\bibitem[{\citenamefont{Reichl et~al.}(2014)\citenamefont{Reichl, Chen, Baer,
  R{\"o}ssler, Ihn, Ensslin, Dietsche, and
  Wegscheider}}]{reichl_increasing_2014}
\bibinfo{author}{\bibfnamefont{C.}~\bibnamefont{Reichl}},
  \bibinfo{author}{\bibfnamefont{J.}~\bibnamefont{Chen}},
  \bibinfo{author}{\bibfnamefont{S.}~\bibnamefont{Baer}},
  \bibinfo{author}{\bibfnamefont{C.}~\bibnamefont{R{\"o}ssler}},
  \bibinfo{author}{\bibfnamefont{T.}~\bibnamefont{Ihn}},
  \bibinfo{author}{\bibfnamefont{K.}~\bibnamefont{Ensslin}},
  \bibinfo{author}{\bibfnamefont{W.}~\bibnamefont{Dietsche}}, \bibnamefont{and}
  \bibinfo{author}{\bibfnamefont{W.}~\bibnamefont{Wegscheider}},
  \bibinfo{journal}{New Journal of Physics} \textbf{\bibinfo{volume}{16}},
  \bibinfo{pages}{023014} (\bibinfo{year}{2014}), ISSN
  \bibinfo{issn}{1367-2630},
  \urlprefix\url{http://iopscience.iop.org/1367-2630/16/2/023014}.

\bibitem[{\citenamefont{Pan et~al.}(2011)\citenamefont{Pan, Masuhara, Sullivan,
  Baldwin, West, Pfeiffer, and Tsui}}]{pan_impact_2011}
\bibinfo{author}{\bibfnamefont{W.}~\bibnamefont{Pan}},
  \bibinfo{author}{\bibfnamefont{N.}~\bibnamefont{Masuhara}},
  \bibinfo{author}{\bibfnamefont{N.~S.} \bibnamefont{Sullivan}},
  \bibinfo{author}{\bibfnamefont{K.~W.} \bibnamefont{Baldwin}},
  \bibinfo{author}{\bibfnamefont{K.~W.} \bibnamefont{West}},
  \bibinfo{author}{\bibfnamefont{L.~N.} \bibnamefont{Pfeiffer}},
  \bibnamefont{and} \bibinfo{author}{\bibfnamefont{D.~C.} \bibnamefont{Tsui}},
  \bibinfo{journal}{Physical Review Letters} \textbf{\bibinfo{volume}{106}},
  \bibinfo{pages}{206806} (\bibinfo{year}{2011}),
  \urlprefix\url{http://link.aps.org/doi/10.1103/PhysRevLett.106.206806}.

\bibitem[{\citenamefont{Gamez and Muraki}(2013)}]{gamez_5/2_2013}
\bibinfo{author}{\bibfnamefont{G.}~\bibnamefont{Gamez}} \bibnamefont{and}
  \bibinfo{author}{\bibfnamefont{K.}~\bibnamefont{Muraki}},
  \bibinfo{journal}{Physical Review B} \textbf{\bibinfo{volume}{88}},
  \bibinfo{pages}{075308} (\bibinfo{year}{2013}),
  \urlprefix\url{http://link.aps.org/doi/10.1103/PhysRevB.88.075308}.

\bibitem[{\citenamefont{Beenakker and van
  Houten}(2004)}]{beenakker_quantum_2004}
\bibinfo{author}{\bibfnamefont{C.~W.~J.} \bibnamefont{Beenakker}}
  \bibnamefont{and} \bibinfo{author}{\bibfnamefont{H.}~\bibnamefont{van
  Houten}}, \bibinfo{journal}{cond-mat/0412664}  (\bibinfo{year}{2004}),
  \bibinfo{note}{\normalfont{Solid State Physics} \textbf{44}, 1 (1991)},
  \urlprefix\url{http://arxiv.org/abs/cond-mat/0412664}.

\bibitem[{\citenamefont{Wharam et~al.}(1988)\citenamefont{Wharam, Thornton,
  Newbury, Pepper, Ahmed, Frost, Hasko, Peacock, Ritchie, and
  Jones}}]{wharam_one-dimensional_1988}
\bibinfo{author}{\bibfnamefont{D.~A.} \bibnamefont{Wharam}},
  \bibinfo{author}{\bibfnamefont{T.~J.} \bibnamefont{Thornton}},
  \bibinfo{author}{\bibfnamefont{R.}~\bibnamefont{Newbury}},
  \bibinfo{author}{\bibfnamefont{M.}~\bibnamefont{Pepper}},
  \bibinfo{author}{\bibfnamefont{H.}~\bibnamefont{Ahmed}},
  \bibinfo{author}{\bibfnamefont{J.~E.~F.} \bibnamefont{Frost}},
  \bibinfo{author}{\bibfnamefont{D.~G.} \bibnamefont{Hasko}},
  \bibinfo{author}{\bibfnamefont{D.~C.} \bibnamefont{Peacock}},
  \bibinfo{author}{\bibfnamefont{D.~A.} \bibnamefont{Ritchie}},
  \bibnamefont{and} \bibinfo{author}{\bibfnamefont{G.~A.~C.}
  \bibnamefont{Jones}}, \bibinfo{journal}{Journal of Physics C: Solid State
  Physics} \textbf{\bibinfo{volume}{21}}, \bibinfo{pages}{L209}
  (\bibinfo{year}{1988}), ISSN \bibinfo{issn}{0022-3719},
  \urlprefix\url{http://iopscience.iop.org/0022-3719/21/8/002}.

\bibitem[{\citenamefont{van Wees et~al.}(1988)\citenamefont{van Wees,
  Kouwenhoven, van Houten, Beenakker, Mooij, Foxon, and
  Harris}}]{van_wees_quantized_1988-1}
\bibinfo{author}{\bibfnamefont{B.~J.} \bibnamefont{van Wees}},
  \bibinfo{author}{\bibfnamefont{L.~P.} \bibnamefont{Kouwenhoven}},
  \bibinfo{author}{\bibfnamefont{H.}~\bibnamefont{van Houten}},
  \bibinfo{author}{\bibfnamefont{C.~W.~J.} \bibnamefont{Beenakker}},
  \bibinfo{author}{\bibfnamefont{J.~E.} \bibnamefont{Mooij}},
  \bibinfo{author}{\bibfnamefont{C.~T.} \bibnamefont{Foxon}}, \bibnamefont{and}
  \bibinfo{author}{\bibfnamefont{J.~J.} \bibnamefont{Harris}},
  \bibinfo{journal}{Physical Review B} \textbf{\bibinfo{volume}{38}},
  \bibinfo{pages}{3625} (\bibinfo{year}{1988}),
  \urlprefix\url{http://link.aps.org/doi/10.1103/PhysRevB.38.3625}.

\bibitem[{\citenamefont{B{\"u}ttiker}(1990)}]{buttiker_quantized_1990}
\bibinfo{author}{\bibfnamefont{M.}~\bibnamefont{B{\"u}ttiker}},
  \bibinfo{journal}{Physical Review B} \textbf{\bibinfo{volume}{41}},
  \bibinfo{pages}{7906} (\bibinfo{year}{1990}),
  \urlprefix\url{http://link.aps.org/doi/10.1103/PhysRevB.41.7906}.

\bibitem[{\citenamefont{Graham et~al.}(2003)\citenamefont{Graham, Thomas,
  Pepper, Cooper, Simmons, and Ritchie}}]{graham_interaction_2003}
\bibinfo{author}{\bibfnamefont{A.~C.} \bibnamefont{Graham}},
  \bibinfo{author}{\bibfnamefont{K.~J.} \bibnamefont{Thomas}},
  \bibinfo{author}{\bibfnamefont{M.}~\bibnamefont{Pepper}},
  \bibinfo{author}{\bibfnamefont{N.~R.} \bibnamefont{Cooper}},
  \bibinfo{author}{\bibfnamefont{M.~Y.} \bibnamefont{Simmons}},
  \bibnamefont{and} \bibinfo{author}{\bibfnamefont{D.~A.}
  \bibnamefont{Ritchie}}, \bibinfo{journal}{Physical Review Letters}
  \textbf{\bibinfo{volume}{91}}, \bibinfo{pages}{136404}
  (\bibinfo{year}{2003}),
  \urlprefix\url{http://link.aps.org/doi/10.1103/PhysRevLett.91.136404}.

\bibitem[{\citenamefont{Ciorga et~al.}(2000)\citenamefont{Ciorga, Sachrajda,
  Hawrylak, Gould, Zawadzki, Jullian, Feng, and
  Wasilewski}}]{ciorga_addition_2000}
\bibinfo{author}{\bibfnamefont{M.}~\bibnamefont{Ciorga}},
  \bibinfo{author}{\bibfnamefont{A.~S.} \bibnamefont{Sachrajda}},
  \bibinfo{author}{\bibfnamefont{P.}~\bibnamefont{Hawrylak}},
  \bibinfo{author}{\bibfnamefont{C.}~\bibnamefont{Gould}},
  \bibinfo{author}{\bibfnamefont{P.}~\bibnamefont{Zawadzki}},
  \bibinfo{author}{\bibfnamefont{S.}~\bibnamefont{Jullian}},
  \bibinfo{author}{\bibfnamefont{Y.}~\bibnamefont{Feng}}, \bibnamefont{and}
  \bibinfo{author}{\bibfnamefont{Z.}~\bibnamefont{Wasilewski}},
  \bibinfo{journal}{Physical Review B} \textbf{\bibinfo{volume}{61}},
  \bibinfo{pages}{R16315} (\bibinfo{year}{2000}),
  \urlprefix\url{http://link.aps.org/doi/10.1103/PhysRevB.61.R16315}.

\bibitem[{\citenamefont{Chklovskii et~al.}(1993)\citenamefont{Chklovskii,
  Matveev, and Shklovskii}}]{chklovskii_ballistic_1993}
\bibinfo{author}{\bibfnamefont{D.~B.} \bibnamefont{Chklovskii}},
  \bibinfo{author}{\bibfnamefont{K.~A.} \bibnamefont{Matveev}},
  \bibnamefont{and} \bibinfo{author}{\bibfnamefont{B.~I.}
  \bibnamefont{Shklovskii}}, \bibinfo{journal}{Physical Review B}
  \textbf{\bibinfo{volume}{47}}, \bibinfo{pages}{12605} (\bibinfo{year}{1993}),
  \urlprefix\url{http://link.aps.org/doi/10.1103/PhysRevB.47.12605}.

\bibitem[{\citenamefont{Sohrmann and
  R{\"o}mer}(2007)}]{sohrmann_compressibility_2007}
\bibinfo{author}{\bibfnamefont{C.}~\bibnamefont{Sohrmann}} \bibnamefont{and}
  \bibinfo{author}{\bibfnamefont{R.~A.} \bibnamefont{R{\"o}mer}},
  \bibinfo{journal}{New Journal of Physics} \textbf{\bibinfo{volume}{9}},
  \bibinfo{pages}{97} (\bibinfo{year}{2007}), ISSN \bibinfo{issn}{1367-2630},
  \urlprefix\url{http://iopscience.iop.org/1367-2630/9/4/097}.

\bibitem[{\citenamefont{Steele}(2006)}]{steele_imaging_2006}
\bibinfo{author}{\bibfnamefont{G.~A.} \bibnamefont{Steele}}, Ph.D. thesis,
  \bibinfo{school}{Massachusetts Institute of Technology}
  (\bibinfo{year}{2006}).

\bibitem[{\citenamefont{{MacDonald}}(1990)}]{macdonald_edge_1990}
\bibinfo{author}{\bibfnamefont{A.~H.} \bibnamefont{{MacDonald}}},
  \bibinfo{journal}{Physical Review Letters} \textbf{\bibinfo{volume}{64}},
  \bibinfo{pages}{220} (\bibinfo{year}{1990}),
  \urlprefix\url{http://link.aps.org/doi/10.1103/PhysRevLett.64.220}.

\bibitem[{\citenamefont{Brey}(1994)}]{brey_edge_1994}
\bibinfo{author}{\bibfnamefont{L.}~\bibnamefont{Brey}},
  \bibinfo{journal}{Physical Review B} \textbf{\bibinfo{volume}{50}},
  \bibinfo{pages}{11861} (\bibinfo{year}{1994}),
  \urlprefix\url{http://link.aps.org/doi/10.1103/PhysRevB.50.11861}.

\bibitem[{\citenamefont{Sim et~al.}(1999)\citenamefont{Sim, Chang, and
  Ihm}}]{sim_composite-fermion_1999}
\bibinfo{author}{\bibfnamefont{H.-S.} \bibnamefont{Sim}},
  \bibinfo{author}{\bibfnamefont{K.~J.} \bibnamefont{Chang}}, \bibnamefont{and}
  \bibinfo{author}{\bibfnamefont{G.}~\bibnamefont{Ihm}},
  \bibinfo{journal}{Physical Review Letters} \textbf{\bibinfo{volume}{82}},
  \bibinfo{pages}{596} (\bibinfo{year}{1999}),
  \urlprefix\url{http://link.aps.org/doi/10.1103/PhysRevLett.82.596}.

\bibitem[{\citenamefont{Halperin}(1982)}]{halperin_quantized_1982}
\bibinfo{author}{\bibfnamefont{B.~I.} \bibnamefont{Halperin}},
  \bibinfo{journal}{Physical Review B} \textbf{\bibinfo{volume}{25}},
  \bibinfo{pages}{2185} (\bibinfo{year}{1982}),
  \urlprefix\url{http://link.aps.org/doi/10.1103/PhysRevB.25.2185}.

\bibitem[{\citenamefont{B{\"u}ttiker}(1988)}]{buttiker_absence_1988}
\bibinfo{author}{\bibfnamefont{M.}~\bibnamefont{B{\"u}ttiker}},
  \bibinfo{journal}{Physical Review B} \textbf{\bibinfo{volume}{38}},
  \bibinfo{pages}{9375} (\bibinfo{year}{1988}),
  \urlprefix\url{http://link.aps.org/doi/10.1103/PhysRevB.38.9375}.

\bibitem[{\citenamefont{Beenakker}(1991)}]{beenakker_theory_1991}
\bibinfo{author}{\bibfnamefont{C.~W.~J.} \bibnamefont{Beenakker}},
  \bibinfo{journal}{Physical Review B} \textbf{\bibinfo{volume}{44}},
  \bibinfo{pages}{1646} (\bibinfo{year}{1991}),
  \urlprefix\url{http://link.aps.org/doi/10.1103/PhysRevB.44.1646}.

\bibitem[{\citenamefont{Beenakker}(1990)}]{beenakker_edge_1990}
\bibinfo{author}{\bibfnamefont{C.~W.~J.} \bibnamefont{Beenakker}},
  \bibinfo{journal}{Physical Review Letters} \textbf{\bibinfo{volume}{64}},
  \bibinfo{pages}{216} (\bibinfo{year}{1990}),
  \urlprefix\url{http://link.aps.org/doi/10.1103/PhysRevLett.64.216}.

\bibitem[{\citenamefont{Jain and Kivelson}(1988)}]{jain_quantum_1988}
\bibinfo{author}{\bibfnamefont{J.~K.} \bibnamefont{Jain}} \bibnamefont{and}
  \bibinfo{author}{\bibfnamefont{S.~A.} \bibnamefont{Kivelson}},
  \bibinfo{journal}{Physical Review Letters} \textbf{\bibinfo{volume}{60}},
  \bibinfo{pages}{1542} (\bibinfo{year}{1988}),
  \urlprefix\url{http://link.aps.org/doi/10.1103/PhysRevLett.60.1542}.

\bibitem[{\citenamefont{van Loosdrecht et~al.}(1988)\citenamefont{van
  Loosdrecht, Beenakker, van Houten, Williamson, van Wees, Mooij, Foxon, and
  Harris}}]{van_loosdrecht_aharonov-bohm_1988}
\bibinfo{author}{\bibfnamefont{P.~H.~M.} \bibnamefont{van Loosdrecht}},
  \bibinfo{author}{\bibfnamefont{C.~W.~J.} \bibnamefont{Beenakker}},
  \bibinfo{author}{\bibfnamefont{H.}~\bibnamefont{van Houten}},
  \bibinfo{author}{\bibfnamefont{J.~G.} \bibnamefont{Williamson}},
  \bibinfo{author}{\bibfnamefont{B.~J.} \bibnamefont{van Wees}},
  \bibinfo{author}{\bibfnamefont{J.~E.} \bibnamefont{Mooij}},
  \bibinfo{author}{\bibfnamefont{C.~T.} \bibnamefont{Foxon}}, \bibnamefont{and}
  \bibinfo{author}{\bibfnamefont{J.~J.} \bibnamefont{Harris}},
  \bibinfo{journal}{Physical Review B} \textbf{\bibinfo{volume}{38}},
  \bibinfo{pages}{10162} (\bibinfo{year}{1988}),
  \urlprefix\url{http://link.aps.org/doi/10.1103/PhysRevB.38.10162}.

\bibitem[{\citenamefont{Venkatachalam et~al.}(2012)\citenamefont{Venkatachalam,
  Hart, Pfeiffer, West, and Yacoby}}]{venkatachalam_local_2012}
\bibinfo{author}{\bibfnamefont{V.}~\bibnamefont{Venkatachalam}},
  \bibinfo{author}{\bibfnamefont{S.}~\bibnamefont{Hart}},
  \bibinfo{author}{\bibfnamefont{L.}~\bibnamefont{Pfeiffer}},
  \bibinfo{author}{\bibfnamefont{K.}~\bibnamefont{West}}, \bibnamefont{and}
  \bibinfo{author}{\bibfnamefont{A.}~\bibnamefont{Yacoby}},
  \bibinfo{journal}{Nature Physics} \textbf{\bibinfo{volume}{8}},
  \bibinfo{pages}{676} (\bibinfo{year}{2012}), ISSN \bibinfo{issn}{1745-2473},
  \urlprefix\url{http://www.nature.com/nphys/journal/v8/n9/full/nphys2384.html}.

\bibitem[{\citenamefont{Girvin}(1984)}]{girvin_particle-hole_1984}
\bibinfo{author}{\bibfnamefont{S.~M.} \bibnamefont{Girvin}},
  \bibinfo{journal}{Physical Review B} \textbf{\bibinfo{volume}{29}},
  \bibinfo{pages}{6012} (\bibinfo{year}{1984}),
  \urlprefix\url{http://link.aps.org/doi/10.1103/PhysRevB.29.6012}.

\bibitem[{\citenamefont{Wen}(1990)}]{wen_electrodynamical_1990}
\bibinfo{author}{\bibfnamefont{X.~G.} \bibnamefont{Wen}},
  \bibinfo{journal}{Physical Review Letters} \textbf{\bibinfo{volume}{64}},
  \bibinfo{pages}{2206} (\bibinfo{year}{1990}),
  \urlprefix\url{http://link.aps.org/doi/10.1103/PhysRevLett.64.2206}.

\bibitem[{\citenamefont{Bid et~al.}(2010)\citenamefont{Bid, Ofek, Inoue,
  Heiblum, Kane, Umansky, and Mahalu}}]{bid_observation_2010}
\bibinfo{author}{\bibfnamefont{A.}~\bibnamefont{Bid}},
  \bibinfo{author}{\bibfnamefont{N.}~\bibnamefont{Ofek}},
  \bibinfo{author}{\bibfnamefont{H.}~\bibnamefont{Inoue}},
  \bibinfo{author}{\bibfnamefont{M.}~\bibnamefont{Heiblum}},
  \bibinfo{author}{\bibfnamefont{C.~L.} \bibnamefont{Kane}},
  \bibinfo{author}{\bibfnamefont{V.}~\bibnamefont{Umansky}}, \bibnamefont{and}
  \bibinfo{author}{\bibfnamefont{D.}~\bibnamefont{Mahalu}},
  \bibinfo{journal}{Nature} \textbf{\bibinfo{volume}{466}},
  \bibinfo{pages}{585} (\bibinfo{year}{2010}), ISSN \bibinfo{issn}{0028-0836},
  \bibinfo{note}{bibtex: Bid2010},
  \urlprefix\url{http://dx.doi.org/10.1038/nature09277}.

\bibitem[{\citenamefont{Wang et~al.}(2013)\citenamefont{Wang, Meir, and
  Gefen}}]{wang_edge_2013-1}
\bibinfo{author}{\bibfnamefont{J.}~\bibnamefont{Wang}},
  \bibinfo{author}{\bibfnamefont{Y.}~\bibnamefont{Meir}}, \bibnamefont{and}
  \bibinfo{author}{\bibfnamefont{Y.}~\bibnamefont{Gefen}},
  \bibinfo{journal}{Physical Review Letters} \textbf{\bibinfo{volume}{111}},
  \bibinfo{pages}{246803} (\bibinfo{year}{2013}),
  \urlprefix\url{http://link.aps.org/doi/10.1103/PhysRevLett.111.246803}.

\bibitem[{\citenamefont{Sarkozy et~al.}(2009)\citenamefont{Sarkozy, Sfigakis,
  Das~Gupta, Farrer, Ritchie, Jones, and Pepper}}]{sarkozy_zero-bias_2009}
\bibinfo{author}{\bibfnamefont{S.}~\bibnamefont{Sarkozy}},
  \bibinfo{author}{\bibfnamefont{F.}~\bibnamefont{Sfigakis}},
  \bibinfo{author}{\bibfnamefont{K.}~\bibnamefont{Das~Gupta}},
  \bibinfo{author}{\bibfnamefont{I.}~\bibnamefont{Farrer}},
  \bibinfo{author}{\bibfnamefont{D.~A.} \bibnamefont{Ritchie}},
  \bibinfo{author}{\bibfnamefont{G.~A.~C.} \bibnamefont{Jones}},
  \bibnamefont{and} \bibinfo{author}{\bibfnamefont{M.}~\bibnamefont{Pepper}},
  \bibinfo{journal}{Physical Review B} \textbf{\bibinfo{volume}{79}},
  \bibinfo{pages}{161307} (\bibinfo{year}{2009}),
  \urlprefix\url{http://link.aps.org/doi/10.1103/PhysRevB.79.161307}.

\bibitem[{\citenamefont{Schnez et~al.}(2011)\citenamefont{Schnez, R{\"o}ssler,
  Ihn, Ensslin, Reichl, and Wegscheider}}]{schnez_imaging_2011}
\bibinfo{author}{\bibfnamefont{S.}~\bibnamefont{Schnez}},
  \bibinfo{author}{\bibfnamefont{C.}~\bibnamefont{R{\"o}ssler}},
  \bibinfo{author}{\bibfnamefont{T.}~\bibnamefont{Ihn}},
  \bibinfo{author}{\bibfnamefont{K.}~\bibnamefont{Ensslin}},
  \bibinfo{author}{\bibfnamefont{C.}~\bibnamefont{Reichl}}, \bibnamefont{and}
  \bibinfo{author}{\bibfnamefont{W.}~\bibnamefont{Wegscheider}},
  \bibinfo{journal}{Physical Review B} \textbf{\bibinfo{volume}{84}},
  \bibinfo{pages}{195322} (\bibinfo{year}{2011}),
  \urlprefix\url{http://link.aps.org/doi/10.1103/PhysRevB.84.195322}.

\bibitem[{\citenamefont{Radu et~al.}(2008)\citenamefont{Radu, Miller, Marcus,
  Kastner, Pfeiffer, and West}}]{radu_quasi-particle_2008}
\bibinfo{author}{\bibfnamefont{I.~P.} \bibnamefont{Radu}},
  \bibinfo{author}{\bibfnamefont{J.~B.} \bibnamefont{Miller}},
  \bibinfo{author}{\bibfnamefont{C.~M.} \bibnamefont{Marcus}},
  \bibinfo{author}{\bibfnamefont{M.~A.} \bibnamefont{Kastner}},
  \bibinfo{author}{\bibfnamefont{L.~N.} \bibnamefont{Pfeiffer}},
  \bibnamefont{and} \bibinfo{author}{\bibfnamefont{K.~W.} \bibnamefont{West}},
  \bibinfo{journal}{Science} \textbf{\bibinfo{volume}{320}},
  \bibinfo{pages}{899 } (\bibinfo{year}{2008}),
  \urlprefix\url{http://www.sciencemag.org/content/320/5878/899.abstract}.

\bibitem[{\citenamefont{Stern and Halperin}(2006)}]{stern_proposed_2006}
\bibinfo{author}{\bibfnamefont{A.}~\bibnamefont{Stern}} \bibnamefont{and}
  \bibinfo{author}{\bibfnamefont{B.~I.} \bibnamefont{Halperin}},
  \bibinfo{journal}{Physical Review Letters} \textbf{\bibinfo{volume}{96}},
  \bibinfo{pages}{016802} (\bibinfo{year}{2006}),
  \urlprefix\url{http://link.aps.org/doi/10.1103/PhysRevLett.96.016802}.

\bibitem[{\citenamefont{Bonderson
  et~al.}(2006{\natexlab{a}})\citenamefont{Bonderson, Kitaev, and
  Shtengel}}]{bonderson_detecting_2006}
\bibinfo{author}{\bibfnamefont{P.}~\bibnamefont{Bonderson}},
  \bibinfo{author}{\bibfnamefont{A.}~\bibnamefont{Kitaev}}, \bibnamefont{and}
  \bibinfo{author}{\bibfnamefont{K.}~\bibnamefont{Shtengel}},
  \bibinfo{journal}{Physical Review Letters} \textbf{\bibinfo{volume}{96}},
  \bibinfo{pages}{016803} (\bibinfo{year}{2006}{\natexlab{a}}),
  \urlprefix\url{http://link.aps.org/doi/10.1103/PhysRevLett.96.016803}.

\bibitem[{\citenamefont{Bonderson
  et~al.}(2006{\natexlab{b}})\citenamefont{Bonderson, Shtengel, and
  Slingerland}}]{bonderson_probing_2006}
\bibinfo{author}{\bibfnamefont{P.}~\bibnamefont{Bonderson}},
  \bibinfo{author}{\bibfnamefont{K.}~\bibnamefont{Shtengel}}, \bibnamefont{and}
  \bibinfo{author}{\bibfnamefont{J.~K.} \bibnamefont{Slingerland}},
  \bibinfo{journal}{Physical Review Letters} \textbf{\bibinfo{volume}{97}},
  \bibinfo{pages}{016401} (\bibinfo{year}{2006}{\natexlab{b}}),
  \urlprefix\url{http://link.aps.org/doi/10.1103/PhysRevLett.97.016401}.

\bibitem[{\citenamefont{Ilan et~al.}(2011)\citenamefont{Ilan, Rosenow, and
  Stern}}]{ilan_signatures_2011}
\bibinfo{author}{\bibfnamefont{R.}~\bibnamefont{Ilan}},
  \bibinfo{author}{\bibfnamefont{B.}~\bibnamefont{Rosenow}}, \bibnamefont{and}
  \bibinfo{author}{\bibfnamefont{A.}~\bibnamefont{Stern}},
  \bibinfo{journal}{Physical Review Letters} \textbf{\bibinfo{volume}{106}},
  \bibinfo{pages}{136801} (\bibinfo{year}{2011}),
  \urlprefix\url{http://link.aps.org/doi/10.1103/PhysRevLett.106.136801}.

\bibitem[{\citenamefont{Ilan et~al.}(2008)\citenamefont{Ilan, Grosfeld, and
  Stern}}]{ilan_coulomb_2008}
\bibinfo{author}{\bibfnamefont{R.}~\bibnamefont{Ilan}},
  \bibinfo{author}{\bibfnamefont{E.}~\bibnamefont{Grosfeld}}, \bibnamefont{and}
  \bibinfo{author}{\bibfnamefont{A.}~\bibnamefont{Stern}},
  \bibinfo{journal}{Physical Review Letters} \textbf{\bibinfo{volume}{100}},
  \bibinfo{pages}{086803} (\bibinfo{year}{2008}),
  \urlprefix\url{http://link.aps.org/doi/10.1103/PhysRevLett.100.086803}.

\bibitem[{\citenamefont{Willett et~al.}(2013)\citenamefont{Willett, Nayak,
  Shtengel, Pfeiffer, and West}}]{willett_magnetic-field-tuned_2013}
\bibinfo{author}{\bibfnamefont{R.~L.} \bibnamefont{Willett}},
  \bibinfo{author}{\bibfnamefont{C.}~\bibnamefont{Nayak}},
  \bibinfo{author}{\bibfnamefont{K.}~\bibnamefont{Shtengel}},
  \bibinfo{author}{\bibfnamefont{L.~N.} \bibnamefont{Pfeiffer}},
  \bibnamefont{and} \bibinfo{author}{\bibfnamefont{K.~W.} \bibnamefont{West}},
  \bibinfo{journal}{Physical Review Letters} \textbf{\bibinfo{volume}{111}},
  \bibinfo{pages}{186401} (\bibinfo{year}{2013}),
  \urlprefix\url{http://link.aps.org/doi/10.1103/PhysRevLett.111.186401}.

\bibitem[{\citenamefont{Bishara et~al.}(2009)\citenamefont{Bishara, Bonderson,
  Nayak, Shtengel, and Slingerland}}]{bishara_interferometric_2009}
\bibinfo{author}{\bibfnamefont{W.}~\bibnamefont{Bishara}},
  \bibinfo{author}{\bibfnamefont{P.}~\bibnamefont{Bonderson}},
  \bibinfo{author}{\bibfnamefont{C.}~\bibnamefont{Nayak}},
  \bibinfo{author}{\bibfnamefont{K.}~\bibnamefont{Shtengel}}, \bibnamefont{and}
  \bibinfo{author}{\bibfnamefont{J.~K.} \bibnamefont{Slingerland}},
  \bibinfo{journal}{Physical Review B} \textbf{\bibinfo{volume}{80}},
  \bibinfo{pages}{155303} (\bibinfo{year}{2009}),
  \urlprefix\url{http://link.aps.org/doi/10.1103/PhysRevB.80.155303}.

\bibitem[{\citenamefont{Miller et~al.}(2007)\citenamefont{Miller, Radu,
  Zumbuhl, Levenson-Falk, Kastner, Marcus, Pfeiffer, and
  West}}]{miller_fractional_2007}
\bibinfo{author}{\bibfnamefont{J.~B.} \bibnamefont{Miller}},
  \bibinfo{author}{\bibfnamefont{I.~P.} \bibnamefont{Radu}},
  \bibinfo{author}{\bibfnamefont{D.~M.} \bibnamefont{Zumbuhl}},
  \bibinfo{author}{\bibfnamefont{E.~M.} \bibnamefont{Levenson-Falk}},
  \bibinfo{author}{\bibfnamefont{M.~A.} \bibnamefont{Kastner}},
  \bibinfo{author}{\bibfnamefont{C.~M.} \bibnamefont{Marcus}},
  \bibinfo{author}{\bibfnamefont{L.~N.} \bibnamefont{Pfeiffer}},
  \bibnamefont{and} \bibinfo{author}{\bibfnamefont{K.~W.} \bibnamefont{West}},
  \bibinfo{journal}{Nat Phys} \textbf{\bibinfo{volume}{3}},
  \bibinfo{pages}{561} (\bibinfo{year}{2007}), ISSN \bibinfo{issn}{1745-2473},
  \urlprefix\url{http://dx.doi.org/10.1038/nphys658}.

\bibitem[{\citenamefont{Wei et~al.}(1985)\citenamefont{Wei, Chang, Tsui, and
  Razeghi}}]{wei_temperature_1985}
\bibinfo{author}{\bibfnamefont{H.~P.} \bibnamefont{Wei}},
  \bibinfo{author}{\bibfnamefont{A.~M.} \bibnamefont{Chang}},
  \bibinfo{author}{\bibfnamefont{D.~C.} \bibnamefont{Tsui}}, \bibnamefont{and}
  \bibinfo{author}{\bibfnamefont{M.}~\bibnamefont{Razeghi}},
  \bibinfo{journal}{Physical Review B} \textbf{\bibinfo{volume}{32}},
  \bibinfo{pages}{7016} (\bibinfo{year}{1985}),
  \urlprefix\url{http://link.aps.org/doi/10.1103/PhysRevB.32.7016}.

\bibitem[{\citenamefont{Siddiki et~al.}(2009)\citenamefont{Siddiki, Horas,
  Moser, Wegscheider, and Ludwig}}]{siddiki_interaction-mediated_2009}
\bibinfo{author}{\bibfnamefont{A.}~\bibnamefont{Siddiki}},
  \bibinfo{author}{\bibfnamefont{J.}~\bibnamefont{Horas}},
  \bibinfo{author}{\bibfnamefont{J.}~\bibnamefont{Moser}},
  \bibinfo{author}{\bibfnamefont{W.}~\bibnamefont{Wegscheider}},
  \bibnamefont{and} \bibinfo{author}{\bibfnamefont{S.}~\bibnamefont{Ludwig}},
  \bibinfo{journal}{{EPL} (Europhysics Letters)} \textbf{\bibinfo{volume}{88}},
  \bibinfo{pages}{17007} (\bibinfo{year}{2009}), ISSN
  \bibinfo{issn}{0295-5075},
  \urlprefix\url{http://iopscience.iop.org/0295-5075/88/1/17007}.

\end{thebibliography}

\end{document}